\documentclass[final,5p,times,twocolumn]{elsarticle}

\usepackage[frozencache,cachedir=.,newfloat]{minted} 

\usepackage{subcaption}
\usepackage[newfloat]{minted} 
\usepackage{hyperref}
\usepackage{tcolorbox}
\usepackage{etoolbox}
\BeforeBeginEnvironment{minted}{\begin{center}\begin{tcolorbox}[width=\linewidth]}
\AfterEndEnvironment{minted}{\end{tcolorbox}\end{center}}
\usepackage{caption}
\usepackage{xspace}
\usepackage[toc,page]{appendix}
\usepackage{amsmath}
\usepackage{algorithm2e}
\RestyleAlgo{ruled}
\SetKwComment{Comment}{\# }{}
\SetKwRepeat{Do}{do}{while}

\newcommand{\inline}[1]{\mintinline{python}{#1}\xspace}
\newcommand{\um}{\mu\mathrm{m}}
\newcommand{\nm}{\mathrm{nm}}
\newcommand{\pyTDGL}{\texttt{pyTDGL} }

\SetupFloatingEnvironment{listing}{name=Code Block}
\BeforeBeginEnvironment{code}{\vspace{10pt}\begin{figure*}[t!]}
\AfterEndEnvironment{code}{\end{figure*}}
\newenvironment{code-onecol}{\captionsetup{type=listing}}{\hfill}
\SetupFloatingEnvironment{listing}{name=Code Block}
\BeforeBeginEnvironment{code-onecol}{\vspace{10pt}\begin{figure}[t!]}
\AfterEndEnvironment{code-onecol}{\end{figure}}

\usepackage{amsmath, amssymb, bm}
\newcounter{bla}

\journal{Computer Physics Communications}

\begin{document}

\begin{frontmatter}



\title{\texttt{pyTDGL}: Time-dependent Ginzburg-Landau in Python}


\author[SIMES,PHYS]{Logan Bishop-Van Horn\corref{lbvh}}\ead{lbvh@stanford.edu}
% \author[PHYS,AP,SIMES]{Kathryn A. Moler}\ead{kmoler@stanford.edu}

\cortext[lbvh]{Corresponding author.}

\address[SIMES]{Stanford Institute for Materials and Energy Sciences, SLAC National Accelerator Laboratory, 2575 Sand Hill Road, Menlo Park, California 94025, USA}
\address[PHYS]{Department of Physics, Stanford University, Stanford, California 94305, USA}
% \address[AP]{Department of Applied Physics, Stanford University, Stanford, California 94305, USA}

\begin{abstract}
Time-dependent Ginzburg-Landau (TDGL) theory is a phenomenological model for the dynamics of superconducting systems. Due to its simplicity in comparison to microscopic theories and its effectiveness in describing the observed properties of the superconducting state, TDGL is widely used to interpret or explain measurements of superconducting devices. Here, we introduce \texttt{pyTDGL}, a Python package that solves a generalized TDGL model for superconducting thin films of arbitrary geometry, enabling simulations of vortex and phase dynamics in mesoscopic superconducting devices. \texttt{pyTDGL} can model the nonlinear magnetic response and dynamics of multiply connected films, films with multiple current bias terminals, and films with a spatially inhomogeneous critical temperature. We demonstrate these capabilities by modeling quasi-equilibrium vortex distributions in irregularly shaped films, and the dynamics and current-voltage-field characteristics of nanoscale superconducting quantum interference devices (nanoSQUIDs).
\end{abstract}


\end{frontmatter}

% \linenumbers

% All CPiP articles must contain the following
% PROGRAM SUMMARY.
\noindent
{\bf PROGRAM SUMMARY}

\begin{small}
\noindent
{\em pyTDGL}\\
% {\em CPC Library link to program files:} (to be added by Technical Editor) \\
{\em Developer's repository link:} \href{http://www.github.com/loganbvh/py-tdgl}{www.github.com/loganbvh/py-tdgl}\\
% {\em Code Ocean capsule:} (to be added by Technical Editor)\\
{\em Licensing provisions:} \href{https://opensource.org/licenses/MIT}{MIT License}\\
{\em Programming language:} Python\\
% {\em Supplementary material:}\\
  % Fill in if necessary, otherwise leave out.
% {\em Journal reference of previous version:}*\\
  %Only required for a New Version summary, otherwise leave out.
% {\em Does the new version supersede the previous version?:}*\\
  %Only required for a New Version summary, otherwise leave out.
% {\em Reasons for the new version:*}\\
  %Only required for a New Version summary, otherwise leave out.
% {\em Summary of revisions:}*\\
  %Only required for a New Version summary, otherwise leave out.
{\em Nature of problem:}  \pyTDGL solves a generalized time-dependent Ginzburg-Landau (TDGL) equation for two-dimensional superconductors of arbitrary geometry, enabling simulations of vortex and phase slip dynamics in thin film superconducting devices.\\
  %Describe the nature of the problem here. \\
{\em Solution method:} The package uses a finite volume adaptive Euler method to solve a coupled TDGL and Poisson equation in two dimensions.\\
  %Describe the method solution here.
% {\em Additional comments including restrictions and unusual features (approx. 50-250 words):}\\
  %Provide any additional comments here.
   \\

% \begin{thebibliography}{0}
% \bibitem{1}Reference 1         % This list should only contain those items referenced in the                 
% \bibitem{2}Reference 2         % Program Summary section.   
% \bibitem{3}Reference 3         % Type references in text as [1], [2], etc.
%                               % This list is different from the bibliography at the end of 
%                               % the Long Write-Up.
% \end{thebibliography}
\end{small}

\tableofcontents

\section{Introduction}
\label{section:introduction}

Ginzburg-Landau (GL) theory~\cite{Ginzburg2008-qb} is a phenomenological theory describing superconductivity in terms of a macroscopic complex order parameter $\psi=|\psi|e^{i\theta}$, which is zero in the normal state and nonzero in the superconducting state. Soon after the introduction of the microscopic Bardeen–Cooper–Schrieffer (BCS) theory~\cite{Bardeen1957-af,Bardeen1957-og}, Gor'kov showed that GL theory can be derived from BCS theory under the assumptions that the temperature $T$ is sufficiently close to the superconducting critical temperature $T_c$ and that the magnetic vector potential $\mathbf{A}(\mathbf{r})$ varies slowly as a function of position $\mathbf{r}$~\cite{Gorkov1959-iv}. As a result, the complex order parameter $\psi(\mathbf{r})$ can be identified as the macroscopic wavefunction of the superconducting condensate and $|\psi(\mathbf{r})|^2$ as the density of Cooper pairs, or the superfluid density. The relevant length scales of the GL model are the coherence length $\xi$, which is the characteristic length for spatial variations in $\psi$, and the London penetration depth $\lambda$, which is the bulk magnetic screening length. The ratio of these length scales, $\kappa=\lambda/\xi$, provides a criterion for the thermodynamic classification of superconductors, where $\kappa<1/\sqrt{2}$ and $\kappa>1/\sqrt{2}$ characterize Type I and Type II superconductors respectively. Both $\xi$ and $\lambda$ are temperature dependent and diverge at $T_c$, but their ratio $\kappa$ is roughly constant over a range of temperatures near $T_c$~\cite{Tinkham2004-ln}.

GL theory describes the equilibrium behavior of superconductors, where the equilibrium value of the order parameter $\psi(\mathbf{r})$ is found by minimizing the Ginzburg-Landau free energy functional. TDGL theory was developed by Gor'kov~\cite{Gorkov1996-do} and Schmid~\cite{Schmid1966-bh} to model the dynamics of the order parameter, however the formal validity of this model is restricted to superconductors in which the gap $\Delta$ vanishes, or those with a large concentration of magnetic impurities. The reason for this restriction, which was pointed out by Gor'kov~\cite{Gorkov1996-do}, is that gapped superconductors exhibit a singularity in the density of states which prohibits approximations that result from expanding quantities in powers of the superconducting gap $\Delta$. This singularity can be broadened by the inclusion of magnetic impurities, restoring the validity of such expansions.

Kramer and Watts-Tobin~\cite{Kramer1978-kb, Watts-Tobin1981-mn} introduced a generalized time-dependent Ginzburg-Landau (gTDGL) model that includes the effect of inelastic electron-phonon scattering, the strength of which is characterized by a parameter $\gamma=2\tau_E\Delta_0$, where $\tau_E$ is the inelastic scattering time and $\Delta_0$ is the zero-field superconducting gap. The extension makes the theory applicable to gapless superconductors ($\gamma=0$) or dirty gapped superconductors ($\gamma > 0$) where the inelastic diffusion length is much smaller than the coherence length $\xi$~\cite{Kopnin2001-ip}.

TDGL and gTDGL have been employed on a wide variety of problems of both fundamental and applied interest~\cite{Aranson2002-so, Blatter1994-mq,Kwok2016-of}. Commercial finite element solvers such as COMSOL have been used to solve the TDGL equations in both two~\cite{Alstrom2011-bc} and three dimensions~\cite{Oripov2020-dq, Peng2017-zt}, and there is an extensive body of literature studying vortex nucleation and dynamics in superconducting devices driven by applied DC~\cite{Machida1993-zm, Clem2011-ji, Clem2012-og, Berdiyorov2012-rn, Sardella2006-hi, Blair2018-og, Jelic2016-ww, Stosic2018-gb, Winiecki2002-ka} and AC~\cite{Hernandez2008-mi, Oripov2020-dq, Bezuglyj2022-cm, Al_Luhaibi2022-cl} electromagnetic fields.
In cases where a commercial solver is not used, the software used to solve the TDGL model may be ``bespoke'' (i.e., tailored to a specific problem and therefore difficult to generalize), closed-source, sparsely documented, and/or require specialized hardware (e.g., graphics processing units, GPUs).
As TDGL is frequently used to explain or interpret measurements of superconducting devices, a portable, open-source, and thoroughly tested and documented software package is desirable to enhance the reproducibility and transferability of research in the field.

\begin{figure}
    \centering
    \includegraphics[width=\linewidth]{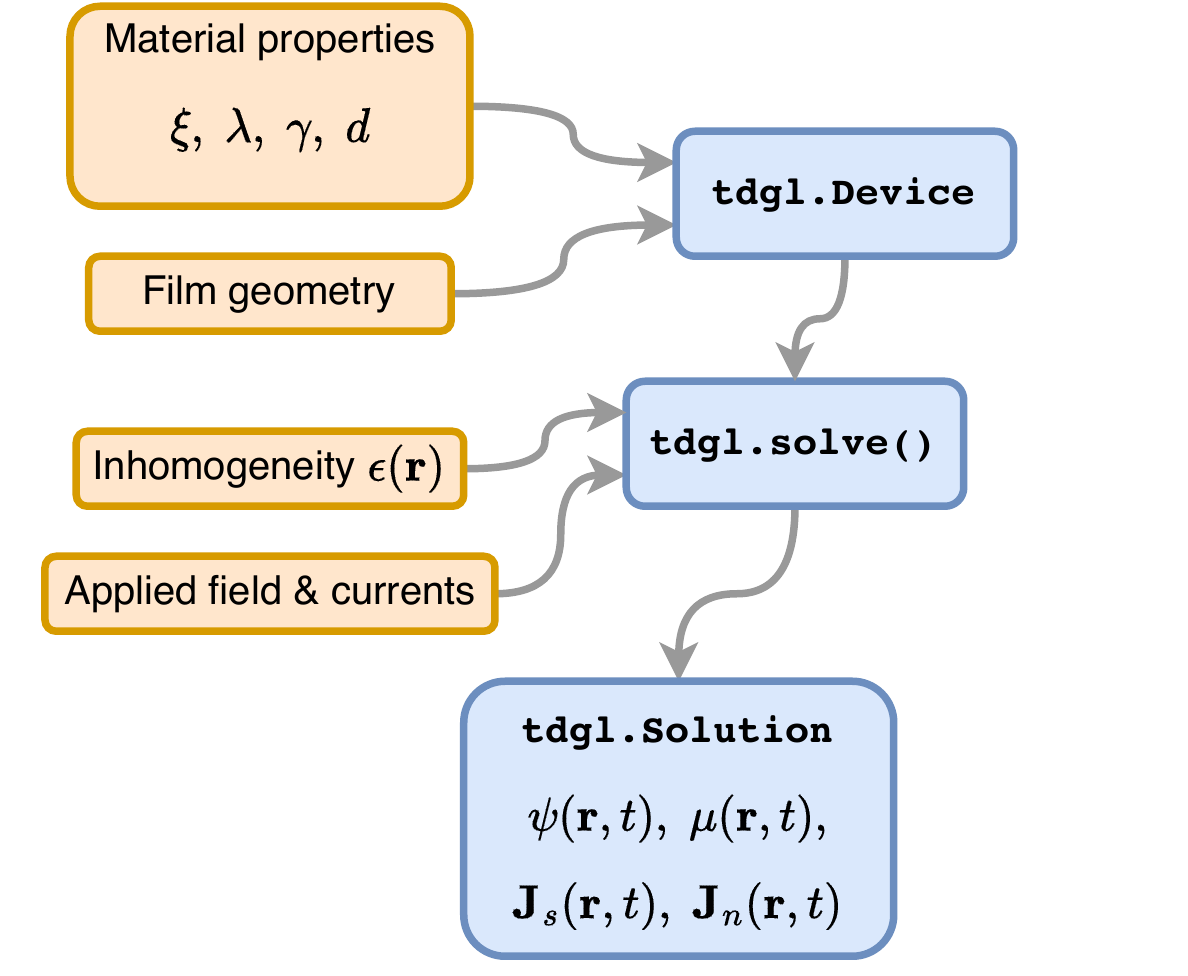}
    \caption{High-level workflow for a \pyTDGL simulation. The material properties and geometry of the superconducting film to be modeled are defined by creating an instance of \inline{tdgl.Device}. A \inline{tdgl.Device} instance, along with the applied magnetic vector potential and transport currents, defines a model that can be solved using \inline{tdgl.solve()}. Calling \inline{tdgl.solve()} returns a \inline{tdgl.Solution} instance, which contains the position- and time-dependent simulation results, along with methods for visualization and post-processing.}
    \label{fig:workflow}
\end{figure}

Here we introduce \texttt{pyTDGL}, an open-source Python package that solves a generalized time-dependent Ginzburg-Landau model in two dimensions, enabling simulations of vortex and phase dynamics in thin film superconducting devices of arbitrary geometry. \texttt{pyTDGL} can model multiply connected films, films with multiple current bias terminals, and films with spatially inhomogeneous critical temperature. The package provides a convenient interface for defining complex device geometries and generating the corresponding finite element data structures, and includes methods for post-processing and visualizing spatially- and temporally-resolved simulation results. A schematic workflow for a \pyTDGL simulation is shown in Figure~\ref{fig:workflow}.

\section{The model}
\label{section:model}

Here we sketch out the generalized time-dependent Ginzburg-Landau model implemented in \texttt{pyTDGL}, and the numerical methods used to solve it. The gTDGL theory is based on Refs. \cite{Kramer1978-kb, Watts-Tobin1981-mn, Jonsson2022-xe, Jonsson2022-mb} and the numerical methods are based on Refs. \cite{Gropp1996-uw, Du1998-kt, Jonsson2022-xe}.

The model is applicable to superconducting thin films of arbitrary geometry. By ``thin'' or ``two-dimensional'' we mean that the film thickness $d$ is much smaller than the coherence length $\xi=\xi(T)$
and the London penetration depth $\lambda=\lambda(T)$, where $T$ is temperature. This assumption implies that both the
superconducting order parameter $\psi(\mathbf{r})$ and the current density $\mathbf{J}(\mathbf{r})$ are roughly
constant over the thickness of the film.
As described above, the model is strictly speaking valid for temperatures very close to the critical
temperature, $T/T_c \to 1$, and for dirty superconductors where the inelastic diffusion length much smaller than the
coherence length $\xi$ \cite{Kramer1978-kb}. Far below
$T_c$, GL theory does not correctly describe the physics of the vortex core, but still accurately captures vortex-vortex interactions~\cite{Aranson2002-so,Kwok2016-of}.

\subsection{Time-dependent Ginzburg-Landau}
\label{section:model-tdgl}

The gTDGL formalism employed here \cite{Kramer1978-kb} consists of a set of coupled partial differential equations for a
complex-valued field $\psi(\mathbf{r}, t)=|\psi(\mathbf{r}, t)|e^{i\theta(\mathbf{r}, t)}$ (the superconducting order parameter) and a real-valued field $\mu(\mathbf{r}, t)$ (the electric scalar potential) which evolve deterministically in time for a given applied magnetic vector potential $\mathbf{A}_\mathrm{applied}(\mathbf{r})$.
By default, we assume that Meissner screening is weak, such that the total vector potential in the film is constant in time and equal to the applied vector potential: $\mathbf{A}(\mathbf{r}, t)=\mathbf{A}_\mathrm{applied}(\mathbf{r})$. The treatment of non-negligible screening is discussed in \ref{appendix:screening}.
% We assume that the magnetic vector potential is either constant in time or varies slowly enough that its time derivative can be neglected when calculating the electric field:
% $\mathbf{E}=-\nabla\mu-\frac{\partial\mathbf{A}}{\partial t}\approx-\nabla\mu$.

In dimensionless units (see Section~\ref{sec:units}), the order parameter $\psi$ evolves according to
\begin{equation}
    \begin{split}
    \frac{u}{\sqrt{1+\gamma^2|\psi|^2}}&\left(\frac{\partial}{\partial t}+i\mu+\frac{\gamma^2}{2}\frac{\partial |\psi|^2}{\partial t}\right)\psi\\
    &= (\epsilon-|\psi|^2)\psi + (\nabla-i\mathbf{A})^2\psi.
    \end{split}
    \label{tdgl}
\end{equation}
The quantity $(\nabla-i\mathbf{A})^2\psi$ is the covariant Laplacian of $\psi$,
which is used in place of an ordinary Laplacian in order to maintain gauge-invariance. Similarly, $(\frac{\partial}{\partial t}+i\mu)\psi$ is the covariant time derivative of $\psi$.
The real-valued parameter $\epsilon(\mathbf{r})=T_c(\mathbf{r})/T - 1 \in [-1,1]$ adjusts the local critical temperature of the film~\cite{Kwok2016-of,Al_Luhaibi2022-cl,Sadovskyy2015-ha}. By default, we set $\epsilon(\mathbf{r})=1$. Setting $\epsilon(\mathbf{r}) < 1$ suppresses the critical temperature at position $\mathbf{r}$, and extended
regions of $\epsilon(\mathbf{r}) < 0$ can be used to model large-scale metallic pinning sites~\cite{Kwok2016-of}. The constant $u=\pi^4/14\zeta(3)\approx5.79$ is the ratio of relaxation times for the amplitude and phase of the order parameter in dirty superconductors
($\zeta(x)$ is the Riemann zeta function), and
$\gamma$ parameterizes the strength of inelastic scattering as described above. Throughout this work, we assume $\gamma=10$.

The electric potential $\mu(\mathbf{r}, t)$ evolves according to the Poisson equation:
\begin{equation}
    \label{poisson}
    \begin{split}
        \nabla^2\mu &= \nabla\cdot\mathrm{Im}[\psi^*(\nabla-i\mathbf{A})\psi]\\
        &=\nabla\cdot\mathbf{J}_s,
    \end{split}
\end{equation}
where $\mathbf{J}_s$ is the dissipationless supercurrent density. The total current density is the sum of the supercurrent density $\mathbf{J}_s$ and the normal current density $\mathbf{J}_n$:
\begin{equation}
    \label{eq:curr-density}
    \begin{split}
    \mathbf{J}&=\mathbf{J}_s+\mathbf{J}_n\\
    &=\mathrm{Im}[\psi^*(\nabla-i\mathbf{A})\psi]-\nabla\mu,
    \end{split}
\end{equation}
where $(\nabla-i\mathbf{A})\psi$ and $\psi^*$
are the covariant gradient and the complex conjugate of $\psi$, respectively. Eq.~\ref{poisson} results from applying the current continuity equation $\nabla\cdot\mathbf{J}=0$ to Eq.~\ref{eq:curr-density}:
\begin{equation}
    \begin{split}
    \nabla\cdot\mathbf{J} &= \nabla\cdot(\mathbf{J}_s+\mathbf{J}_n)\\
    &=\nabla\cdot\mathrm{Im}[\psi^*(\nabla-i\mathbf{A})\psi]-\nabla^2\mu\\
    &=0.
    \end{split}
\end{equation}
For thin films ($d\ll\lambda$), the thickness-integrated sheet current density is
\begin{equation}
    \mathbf{K} = \mathbf{K}_s + \mathbf{K}_n = d(\mathbf{J}_s+\mathbf{J}_n).
\end{equation}

In addition to the electric potential (Eq.~\ref{poisson}), one can couple the dynamics of the order parameter (Eq.~\ref{tdgl}) to other physical quantities to create a ``multiphysics'' model. For example, it is common to couple the TDGL equations to the local temperature $T(\mathbf{r}, t)$ of the superconductor via a heat balance equation to model self-heating~\cite{Gurevich1987-sv, Berdiyorov2012-rn, Zotova2012-nc, Jelic2016-ww, Jing2018-qc}.

\subsection{Boundary conditions}
\label{sec:boundary}
Isolating boundary conditions are enforced on superconductor-vacuum interfaces, in the form of Neumann boundary conditions for $\psi$ and $\mu$:
\begin{subequations}
    \label{eq:bc-vacuum}
    \begin{align}
        \hat{\mathbf{n}}\cdot(\nabla-i\mathbf{A})\psi &= 0\label{eq:bc-vacuum-psi}\\
        \hat{\mathbf{n}}\cdot\nabla\mu &= 0\label{eq:bc-vacuum-mu},
    \end{align}
\end{subequations}
where $\hat{\mathbf{n}}$ is a unit vector normal to the interface.
Superconductor-normal metal interfaces can be used to apply a bias current density $J_\mathrm{ext}$.
For such interfaces, we impose Dirichlet boundary conditions on $\psi$ and Neumann boundary conditions on $\mu$:
\begin{subequations}
    \label{eq:bc-normal}
    \begin{align}
        \psi &= 0 \label{eq:bc-normal-psi}\\
        \hat{\mathbf{n}}\cdot\nabla\mu &= J_\mathrm{ext}\label{eq:bc-normal-mu}.
    \end{align}
\end{subequations}
A film can have an arbitrary number of normal metal current terminals.
If we label the terminals $k=1,2,\ldots$, we can express the global current conservation constraint as
\begin{equation}
    \label{current-cons}
    \sum_k I_{\mathrm{ext},k} = \sum_k J_{\mathrm{ext},k}L_k = 0,  
\end{equation}
where $I_{\mathrm{ext},k}$ is the total current through terminal $k$, $L_k$ is the length of terminal $k$,
and $J_{\mathrm{ext},k}$ is the current density along terminal $k$, which we assume to be uniform in magnitude along the terminal and directed normal to the terminal.
From Eq.~\ref{current-cons}, it follows that the current boundary condition for terminal $k$ is:
\begin{equation}
    \label{eq:bc-current}
    J_{\mathrm{ext},k}=-\frac{1}{L_k}\sum_{\ell\neq k}I_{\mathrm{ext},\ell}=-\frac{1}{L_k}\sum_{\ell\neq k}J_{\mathrm{ext},\ell}L_\ell.   
\end{equation}

\begin{figure*}[h]
    \centering
    \includegraphics[width=\textwidth]{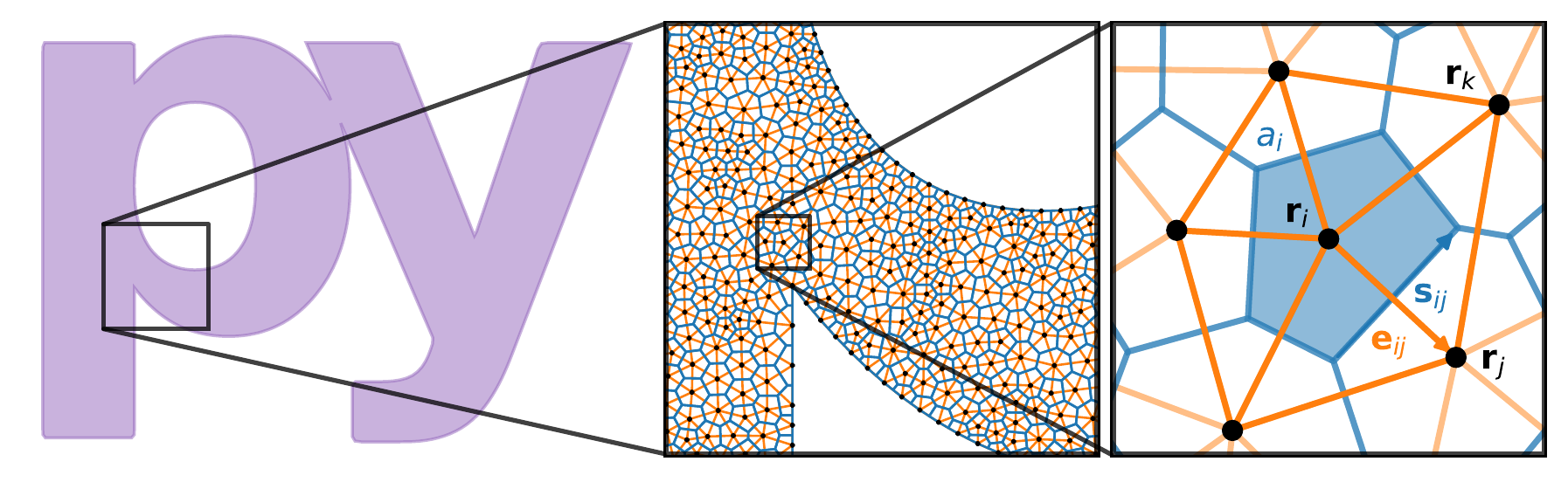}
    \caption{The structure of a finite volume mesh. {\bf Left}: A superconducting film in the shape of the letters ``py''. {\bf Center}: A portion of the finite volume mesh for the film. The mesh sites are shown as black dots, and the triangular Delaunay mesh and dual Voronoi mesh are shown in orange and blue respectively. {\bf Right}: The mesh site $\mathbf{r}_i$ is connected to its nearest neighbors $\mathcal{N}(i)$ by edges $\mathbf{e}_{ij}=\mathbf{r}_j-\mathbf{r}_i$. Edges (orange lines) that connect three adjacent sites ($i, j, k$) in a counterclockwise fashion form a triangular element $c_{ijk}$ of the Delaunay mesh. The Voronoi region surrounding site $i$ (blue) has an area $a_i$, and the edge of the Voronoi cell that intersects edge $\mathbf{e}_{ij}$ has a length $s_{ij}=|\mathbf{s}_{ij}|$. The indexing of mesh sites is arbitrary but fixed.}
    \label{fig:voronoi}
\end{figure*}

\subsection{Units}
\label{sec:units}
The TDGL equations (Eqs.~\ref{tdgl} and \ref{poisson}) are solved in dimensionless units, where $\psi$ is measured in units of $|\psi_0|$, the magnitude of the order parameter in the absence of applied fields or currents. In these units, we have $0 \leq |\psi|^2=n_s \leq 1$, where $n_s$ is the superfluid density normalized to its zero-field value. The scale factors used to convert between physical and dimensionless units are given in terms of material parameters and fundamental constants,
namely the superconducting coherence length $\xi$, London penetration depth $\lambda$, normal state conductivity $\sigma$, film thickness $d$, vacuum permeability $\mu_0$, and superconducting flux quantum $\Phi_0=h/2e$, where $h$ is Planck's constant and $e>0$ is the elementary charge. 

Distance is measured in units of $\xi$.
Time is measured in units of
$$\tau_0 = \mu_0\sigma\lambda^2.$$
Magnetic field is measured in units of the upper critical field
$$B_0=B_{c2}=\mu_0H_{c2} = \frac{\Phi_0}{2\pi\xi^2}.$$
Magnetic vector potential is measured in units of
$$A_0=\xi B_0=\frac{\Phi_0}{2\pi\xi}.$$
Current density is measured in units of
$$J_0=\frac{4\xi B_{c2}}{\mu_0\lambda^2}.$$
Sheet current density is measured in units of
$$K_0=J_0 d=\frac{4\xi B_{c2}}{\mu_0\Lambda},$$
where $\Lambda=\lambda^2/d$ is the effective magnetic penetration depth.
The electric potential $\mu$ is measured in units of 
$$V_0=\xi J_0/\sigma=\frac{4\xi^2 B_{c2}}{\mu_0\sigma\lambda^2}.$$

\section{Numerical implementation}
\label{sec:numerics}

\subsection{Finite volume method}
\label{sec:finite-volume}

To simulate superconducting films of any shape, we solve the TDGL model on an unstructured Delaunay mesh in two dimensions \cite{Du1998-kt, Jonsson2022-xe}, as illustrated in Figure~\ref{fig:voronoi}. The mesh is composed of a set of sites, each denoted by its position $\mathbf{r}_i\in\mathbb{R}^2$ or an integer index $i$,
and a set of triangular cells $c_{ijk}$. Nearest-neighbor sites $\mathbf{r}_i$ and $\mathbf{r}_j$ are connected by an edge, denoted by the vector $\mathbf{e}_{ij}=\mathbf{r}_j-\mathbf{r}_i$ or the 2-tuple $(i, j)$, which
has a length $e_{ij}=|\mathbf{e}_{ij}|$ and a direction $\hat{\mathbf{e}}_{ij}=\mathbf{e}_{ij}/e_{ij}$. Each cell $c_{ijk}=(i, j, k)$ represents a triangle with three edges
($(i, j)$, $(j, k)$, and $(k, i)$) that connect sites $\mathbf{r}_i$, $\mathbf{r}_j$, $\mathbf{r}_k$ in
a counterclockwise fashion. Each site is assigned an effective area $a_i$, which is the area of the Voronoi cell surrounding the site. The Voronoi cell surrounding site $i$ consists of all points in $\mathbb{R}^2$ that are closer to site $\mathbf{r}_i$ than to any other site in the mesh. The side of the Voronoi cell that intersects edge $(i, j)$ is denoted $\mathbf{s}_{ij}$ and has a length $s_{ij}$. The collection of all Voronoi cells tessellates the film and forms a mesh that is dual to the triangular Delaunay mesh.

A scalar function $f(\mathbf{r}, t)$ can be discretized at a given time $t^{n}$
as the value of the function on each site, $f_i^{n}=f(\mathbf{r}_i, t^{n})$.
A vector function $\mathbf{F}(\mathbf{r}, t)$ can be discretized at time $t^{n}$ as the flow of the vector field between sites, $F_{ij}^{n}=\mathbf{F}((\mathbf{r}_i+\mathbf{r}_j)/2, t^{n})\cdot\hat{\mathbf{e}}_{ij}$, where $(\mathbf{r}_i+\mathbf{r}_j)/2=\mathbf{r}_{ij}$
is the center of edge $(i, j)$. The gradient of a scalar function $g(\mathbf{r})$ is approximated on the edges of the mesh. The value of $\nabla g$
at position $\mathbf{r}_{ij}$ is:
\begin{equation}
    \label{gradient}
    (\nabla g)_{ij}=\left.(\nabla g)\right|_{\mathbf{r}_{ij}}\approx\frac{g_j-g_i}{e_{ij}}\hat{\mathbf{e}}_{ij}.
\end{equation}
To calculate the divergence of a vector field $\mathbf{F}(\mathbf{r})$ on the mesh, we assume that each Voronoi cell is small enough that the value of $\nabla\cdot\mathbf{F}$ is constant over the area of the cell and equal to the value at the mesh site lying inside the cell, $\mathbf{r}_i$.
Then, using the divergence theorem in two dimensions, we have
\begin{equation}
    \label{divergence}
    \begin{split}
        \int(\nabla\cdot\mathbf{F})\,\mathrm{d}^2\mathbf{r} &= \oint(\mathbf{F}\cdot\hat{\mathbf{n}})\,\mathrm{d}s\\
        \left.(\nabla\cdot\mathbf{F})a_i\right|_{\mathbf{r}_i}&\approx\sum_{j\in\mathcal{N}(i)}F_{ij}s_{ij}\\
        (\nabla\cdot\mathbf{F})_i=\left.(\nabla\cdot\mathbf{F})\right|_{\mathbf{r}_i}&\approx\frac{1}{a_i}\sum_{j\in\mathcal{N}(i)}F_{ij}s_{ij},
    \end{split}
\end{equation}
where $\mathcal{N}(i)$ is the set of sites adjacent to site $\mathbf{r}_i$. The Laplacian of a scalar function $g$ is given by $\nabla^2 g=\nabla\cdot\nabla g$, so combining Eqs.~\ref{gradient} and \ref{divergence} we have
\begin{equation}
    \label{laplacian}
    (\nabla^2g)_i=\left.(\nabla^2 g)\right|_{\mathbf{r}_i}\approx\frac{1}{a_i}\sum_{j\in\mathcal{N}(i)}\frac{g_j-g_i}{e_{ij}}s_{ij}.
\end{equation}

The discrete gradient, divergence, and Laplacian of a field at site $i$ depend only on the value of the field at
site $i$ and its nearest neighbors. This means that the corresponding operators, Eqs.~\ref{gradient}, \ref{divergence}, and \ref{laplacian},
can be efficiently represented as sparse matrices, and their action given by sparse matrix-vector products.

\subsection{Covariant derivatives}
\label{sec:covariant}

We use link variables to construct covariant versions of the spatial derivatives and time derivatives of $\psi$~\cite{Gropp1996-uw, Du1998-kt, Du1998-rb}. In the discrete case corresponding to our finite volume method, this amounts to adding a complex phase whenever taking a difference
in $\psi$ between mesh sites (for spatial derivatives) or time steps (for time derivatives).

The covariant gradient of $\psi$ at time $t^{n}$ and edge $\mathbf{r}_{ij}$ is:
\begin{equation}
    \label{grad-psi}
    \left.\left(\nabla-i\mathbf{A}\right)\psi\right|_{\mathbf{r}_{ij}}^{t^{n}}=\frac{U^{n}_{ij}\psi_j^{n}-\psi_i^{n}}{e_{ij}}\hat{\mathbf{e}}_{ij},  
\end{equation}
where $U^{n}_{ij}=\exp(-i\mathbf{A}(\mathbf{r}_{ij}, t^{n})\cdot\mathbf{e}_{ij})$ is the spatial link variable.
Eq.~\ref{grad-psi} is similar to the gauge-invariant phase difference
in Josephson junction physics. The covariant Laplacian of $\psi$ at time $t^{n}$ and site $\mathbf{r}_i$ is:
\begin{equation}
    \label{lapacian-psi}
    \left.\left(\nabla-i\mathbf{A}\right)^2\psi\right|_{\mathbf{r}_{i}}^{t^{n}}=\frac{1}{a_i}\sum_{j\in\mathcal{N}(i)}\frac{U^{n}_{ij}\psi_j^{n}-\psi_i^{n}}{e_{ij}}s_{ij}  
\end{equation}
The covariant time derivative of $\psi$ at time $t^{n}$ and site $\mathbf{r}_i$ is
\begin{equation}
    \label{dpsi_dt}
    \left.\left(\frac{\partial}{\partial t}+i\mu\right)\psi\right|_{\mathbf{r}_i}^{t^{n}}=\frac{U_i^{n, n+1}\psi_i^{n+1}-\psi_i^{n}}{\Delta t^{n}},
\end{equation}
where $U_i^{n, n+1}=\exp(i\mu_i^{n}\Delta t^{n})$ is the temporal link variable.

\subsection{Euler method}
\label{sec:euler}

The discretized form of the equations of motion for $\psi(\mathbf{r}, t)$ (Eq.~\ref{tdgl}) and $\mu(\mathbf{r}, t)$ (Eq.~\ref{poisson}) are:
\begin{equation}
    \label{tdgl-num2}
    \begin{split}
        &\frac{u}{\Delta t^{n}\sqrt{1 + \gamma^2\left|\psi_i^{n}\right|^2}}
        \left[
            \psi_i^{n+1}e^{i\mu_i^{n}\Delta t^{n}}-\psi_i^{n}
            +\frac{\gamma^2}{2}\left(\left|\psi_i^{n+1}\right|^2-\left|\psi_i^{n}\right|^2\right)\psi_i^{n}
        \right]\\
        &\quad=\left(\epsilon_i-\left|\psi_i^{n}\right|^2\right)\psi_i^{n}
        +\frac{1}{a_i}\sum_{j\in\mathcal{N}(i)}\frac{U^{n}_{ij}\psi_j^{n}-\psi_i^{n}}{e_{ij}}s_{ij}
    \end{split}
\end{equation}
and
\begin{equation}
    \label{poisson-num}
    \begin{split}
    \sum_{j\in\mathcal{N}(i)}\frac{\mu_j^{n}-\mu_i^{n}}{e_{ij}}s_{ij}&=\sum_{j\in\mathcal{N}(i)}J_{s,ij}^{n}|s_{ij}|
    \end{split}
\end{equation}
respectively, where supercurrent $J_{ij}^{n}$ is given by
\begin{equation}
    \label{eq:supercurrent}
    J^{n}_{s,ij}=\mathrm{Im}\left\{\left(\psi_i^{n}\right)^*\,\frac{U^{n}_{ij}\psi_j^{n}-\psi_i^{n}}{e_{ij}}\right\}.
\end{equation}
The superscript $n$ is an integer index denoting the solve step or iteration and $\Delta t^n$ is the time step in units of $\tau_0$. If we isolate the terms in Eq.~\ref{tdgl-num2} involving the order parameter at time $t^{n+1}$, we can rewrite Eq.~\ref{tdgl-num2} in the form
\begin{equation}
    \label{quad-1}
    \psi_i^{n+1}+z_i^{n}\left|\psi_i^{n+1}\right|^2=w_i^{n},
\end{equation}
where 
\begin{equation}
    \label{eq:z}
    z_i^{n}=\frac{\gamma^2}{2}e^{-i\mu_i^{n}\Delta t^{n}}\psi_i^{n}
\end{equation}
and
\begin{equation}
    \label{eq:w}
    \begin{split}
    w_i^{n}=&z_{i}^{n}\left|\psi_i^{n}\right|^2+e^{-i\mu_i^{n}\Delta t^{n}}\,\times\\
    &\Biggl[\psi_i^{n}+\frac{\Delta t^{n}}{u}\sqrt{1+\gamma^2\left|\psi_i^{n}\right|^2}\,\times\\
    &\quad\biggl(
        \left(\epsilon_i-\left|\psi_i^{n}\right|^2\right)\psi_{i}^{n} +
        \frac{1}{a_i}\sum_{j\in\mathcal{N}(i)}\frac{U^{n}_{ij}\psi_j^{n}-\psi_i^{n}}{e_{ij}}s_{ij}
    \biggr)
    \Biggr].
    \end{split}
\end{equation}
Solving Eq.~\ref{quad-1} for $\left|\psi_i^{n+1}\right|^2$,
we arrive at a quadratic equation in $\left|\psi_i^{n+1}\right|^2$:
% \begin{equation}
%     \label{eq:quad-2}
%     \begin{split}
%     0 =& \left|z_i^{n}\right|^2\left|\psi_i^{n+1}\right|^4\\
%     &-\left(2\left[
%         \mathrm{Re}\left\{z_i^{n}\right\}\mathrm{Re}\left\{w_i^{n}\right\}
%         +\mathrm{Im}\left\{z_i^{n}\right\}\mathrm{Im}\left\{w_i^{n}\right\}
%     \right] + 1\right)\left|\psi_i^{n+1}\right|^2\\
%     &+ \left|w_i^{n}\right|^2,
%     \end{split}
% \end{equation}
\begin{equation}
    \label{eq:quad-2}
    \left|z_i^{n}\right|^2\left|\psi_i^{n+1}\right|^4 - (2c_i^n+1)\left|\psi_i^{n+1}\right|^2+\left|w_i^{n}\right|^2=0,
\end{equation}
where we have defined 
\begin{equation}
    \label{eq:c}
    c_i^{n}=
    \mathrm{Re}\left\{z_i^{n}\right\}\mathrm{Re}\left\{w_i^{n}\right\}
    +\mathrm{Im}\left\{z_i^{n}\right\}\mathrm{Im}\left\{w_i^{n}\right\}.
\end{equation}
See~\ref{appendix:euler} for an explicit derivation of Eq.~\ref{eq:quad-2}.
To solve Eq.~\ref{eq:quad-2}, which has the form $ax^2+bx+c=0$, we use a modified quadratic formula:
%\footnote{This equation is sometimes called the ``citardauq'' formula, which is ``quadratic'' spelled backwards.}:
\begin{equation}
    \label{eq:citardauq}
    \begin{split}
        x &= \frac{-b\pm\sqrt{b^2-4ac}}{2a}\cdot\frac{-b\mp\sqrt{b^2-4ac}}{-b\mp\sqrt{b^2-4ac}}\\
        % &=\frac{b^2-(b^2-4ac)}{2a(-b\mp\sqrt{b^2-4ac})}\\
        % &=\frac{4ac}{2a(-b\mp\sqrt{b^2-4ac})}\\
        &=\frac{2c}{-b\mp\sqrt{b^2-4ac}}
    \end{split}
\end{equation}
in order to avoid numerical issues when $a=\left|z_i^n\right|^2=0$, i.e., when $\left|\psi_i^n\right|^2=0$ or $\gamma=0$.
Applying Eq.~\ref{eq:citardauq} to Eq.~\ref{quad-1} yields
\begin{equation}
    \label{eq:quad-root}
    \left|\psi_i^{n+1}\right|^2=\frac{2\left|w_i^{n}\right|^2}{(2c_i^{n} + 1)+\sqrt{(2c_i^{n} + 1)^2 - 4\left|z_i^{n}\right|^2\left|w_i^{n}\right|^2}}.
\end{equation}
We take the root with the ``$+$'' sign in Eq.~\ref{eq:quad-root} because the ``$-$'' sign results in unphysical behavior where
$\left|\psi_i^{n+1}\right|^2$ diverges when $\left|z_i^{n}\right|^2$ vanishes (i.e., when $\left|\psi_i^{n}\right|^2$ and/or $\gamma$ is zero).

Combining Eq.~\ref{quad-1} and Eq.~\ref{eq:quad-root} allows us to find the order parameter at time $t^{n+1}$ in terms of the 
order parameter and the electric potential at time $t^{n}$:
\begin{equation}
    \label{eq:psi-sol}
    \begin{split}
    \psi_i^{n+1} &= w_i^{n} - z_i^{n}\left|\psi_i^{n+1}\right|^2\\
    &=w_i^{n} - z_i^{n}\frac{2\left|w_i^{n}\right|^2}{(2c_i^{n} + 1)+\sqrt{(2c_i^{n} + 1)^2 - 4\left|z_i^{n}\right|^2\left|w_i^{n}\right|^2}}.
    \end{split}
\end{equation}
Combining Eq.~\ref{eq:psi-sol} and Eq.~\ref{poisson-num} yields a sparse linear system that can be solved to find
$\mu_i^{n+1}$ given $\mu_i^{n}$ and $\psi_i^{n + 1}$. Eq.~\ref{poisson-num} is solved using sparse LU decomposition~\cite{Li2005-gv}.

\subsection{Adaptive time step}
\label{sec:adaptive}

\pyTDGL uses an adaptive time step algorithm that adjusts the time step $\Delta t^{n}$
based on the speed of the system's dynamics, which can significantly reduce the wall-clock time of a simulation without sacrificing solution accuracy (See~\ref{appendix:adpative}). There are three parameters that control the adaptive time step algorithm,
$\Delta t_\mathrm{init}$, $\Delta t_\mathrm{max}$, and $N_\mathrm{window}$. The initial time step in iteration $n=0$ is set to $\Delta t^{0}=\Delta t_\mathrm{init}$. We keep a running list of
$\Delta|\psi|^2_n=\max_i \left|\left(\left|\psi_i^{n}\right|^2-\left|\psi_i^{n-1}\right|^2\right)\right|$ for each iteration $n$. Then, for each iteration $n > N_\mathrm{window}$, we define a tentative new time step $\Delta t_\star$
using the following heuristic:
\begin{subequations}
    \label{eq:dt-tentative}
    \begin{align}
        \delta_n &= \frac{1}{N_\mathrm{window}}\sum_{\ell=0}^{N_\mathrm{window}-1}\Delta|\psi|^2_{n-\ell}\\
        \Delta t_\star & = \min\left(\frac{1}{2}\left(\Delta t^n +  \frac{\Delta t_\mathrm{init}}{\delta_n}\right),\;\Delta t_\mathrm{max}\right)
    \end{align}
\end{subequations}

Eq.~\ref{eq:dt-tentative} has the effect of automatically selecting a small time step if the recent dynamics of the order parameter are fast (i.e., if $\delta_n$ is large) and a larger time step if the dynamics are slow.\footnote{Because new time steps are chosen based on the dynamics of the magnitude of the order parameter, we recommend disabling
the adaptive time step algorithm or using a strict $\Delta t_\mathrm{max}$ in cases where the entire
film is in the normal state, $\psi=0$.} The the tentative time step selected at solve step $n$ may be too large to accurately solve for the state of the system in step $m=n+1$. We detect such a failure to converge by evaluating the discriminant of
Eq.~\ref{eq:quad-2}. If the discriminant, $(2c_i^{m} + 1)^2 - 4|z_i^{m}|^2|w_i^{m}|^2$, is less than zero for any
site $i$, then the value of $|\psi_i^{m+1}|^2$ found in Eq.~\ref{eq:quad-root} will be complex, which is unphysical.
If this happens, we iteratively reduce the time step $\Delta t^{m}$ (setting $\Delta t^{m} \leftarrow \Delta t^{m}\times M_\mathrm{adaptive}$ at each iteration, where $0<M_\mathrm{adaptive}<1$ is a user-defined multiplier) and re-solve Eq.~\ref{eq:quad-2} until
the discriminant is nonnegative for all sites $i$, then proceed with the rest of the calculation for step $m$. If this process fails to find a suitable $\Delta t^m$ in $N_\mathrm{retries}^\mathrm{max}$ iterations, where $N_\mathrm{retries}^\mathrm{max}$ is a parameter that can be set by the user, the solver raises an error. Pseudocode for the main solver is given in Algorithms~\ref{alg:adaptive-euler-step} and \ref{alg:adaptive} in \ref{appendix:pseudocode}.

\section{Package overview}
\label{sec:package-overview}
In this section, we provide an overview of the \texttt{pyTDGL} Python package. The package is written for and tested on Python versions 3.8, 3.9, and 3.10. The numerical methods described in Section~\ref{sec:numerics} are implemented using NumPy~\cite{Harris2020-xv} and SciPy~\cite{Virtanen2020-zz}, data visualization is implemented using Matplotlib~\cite{Hunter2007-il}, and data storage is implemented using the HDF5 file format via the h5py library~\cite{Collette2013-rq}. Physical units are managed using Pint~\cite{Grecco}, and finite element meshes are generated using the MeshPy~\cite{Klockner} Python interface to Triangle~\cite{Shewchuk1996-va}, a fast compiled 2D mesh generator.
%JAX~\cite{jax2018github} is an optional dependency that is recommended for users who would like to model screening, as it enables GPU acceleration (see \ref{appendix:screening}).
The \pyTDGL Python user interface is adapted from SuperScreen~\cite{Bishop-Van_Horn2022-sy}, and some of the finite volume and interactive visualization code is adapted from SuperDetectorPy\footnote{\href{https://github.com/afsa/super-detector-py}{https://github.com/afsa/super-detector-py}}, a public GitHub repository accompanying Refs.~\cite{Jonsson2022-xe,Jonsson2022-mb}.

\pyTDGL is hosted on GitHub,\footnote{\href{https://github.com/loganbvh/py-tdgl}{https://github.com/loganbvh/py-tdgl}} and automated testing is performed using the GitHub Actions continuous integration (CI) tool. The package can be installed from GitHub or PyPI, the Python Package Index.\footnote{\href{https://pypi.org/project/tdgl/}{https://pypi.org/project/tdgl/}} The source code and online documentation\footnote{\href{https://py-tdgl.readthedocs.io/}{https://py-tdgl.readthedocs.io/}} are provided under the MIT License.\footnote{\href{https://opensource.org/licenses/MIT}{https://opensource.org/licenses/MIT}} The stable version of the package corresponding to this manuscript is \texttt{v0.2.1}.
% A schematic workflow for a \pyTDGL simulation is shown in Figure~\ref{fig:workflow}.

\subsection{Device interface}
\label{sec:device}

The material properties and geometry of a superconducting device to be modeled are defined by creating an instance of the \inline{tdgl.Device} class, which is responsible for generating the finite element mesh and translating between physical units (e.g., millitesla and microns) and the dimensionless units described in Section~\ref{sec:units}. A \inline{tdgl.Device} consists of a \inline{tdgl.Layer} and one or more \inline{tdgl.Polygons}. The \inline{Layer} stores the coherence length $\xi$, London penetration depth $\lambda$, inelastic scattering parameter $\gamma$, and film thickness $d$. Instances of \inline{tdgl.Polygon} are used to define the shape of the superconducting film, including any holes in the film, and to define the location of any normal metal current terminals. \inline{Polygons} can be manipulated using affine transformations (translation, scaling, rotation) and combined using constructive solid geometry methods (see Table~\ref{table:polygon}) to generate complex shapes from simple geometric primitives such as rectangles and ellipses . One can also define any number of ``probe points,'' which are positions $\mathbf{r}_{p,i}$ in the film for which the electric potential  $\mu(\mathbf{r}_{p,i}, t^n)$ and phase $\theta(\mathbf{r}_{p,i}, t^n)=\arg\psi(\mathbf{r}_{p,i}, t^n)$ are measured at each solve step $n$.

\begin{table}[h]
\centering
\begin{tabular}{|c|c|c|}
\hline
\inline{tdgl.Polygon} method & Python operator & Set operation\\ \hline \hline
\inline{A.union(B)} & \inline{A + B}        & $A\cup B$ \\
\inline{A.intersection(B)} & \inline{A * B} & $A\cap B$ \\
\inline{A.difference(B)} & \inline{A - B}  & $A\setminus B$\\
\hline
\end{tabular}
\caption{Constructive solid geometry methods for combining \inline{tdgl.Polygon} objects, \inline{A} and \inline{B}, along with their corresponding set-theoretic operations. The object \inline{B} can be of any type that can be converted to a \inline{tdgl.Polygon}, e.g., a Numpy array of polygon vertices of shape \inline{(n, 2)}. The set $A$ ($B$) denotes all points in $\mathbb{R}^2$ that are inside the polygon \inline{A} (\inline{B}).}
\label{table:polygon}
\end{table}

Once the geometry has been defined, the finite volume mesh (Figure~\ref{fig:voronoi}) can be generated by calling \inline{tdgl.Device.make_mesh()}. To control the resolution of the mesh, one can specify the minimum number of sites allowed in the mesh (e.g., \inline{min_points = 5000}) and/or the maximum allowed edge length (e.g., \inline{max_edge_length = xi / 2}, where \inline{xi} is the coherence length). The lengths of the edges in the mesh should be small compared to the coherence length to ensure solution accuracy. The device can be inspected visually by calling \inline{tdgl.Device.plot()} or \inline{tdgl.Device.draw()} (see Figure~\ref{fig:squid-schematic}), and information about the mesh can be displayed using \inline{tdgl.Device.mesh_stats()}. A device can be saved to disk using \inline{tdgl.Device.to_hdf5()} and loaded from disk using \inline{tdgl.Device.from_hdf5()}.

\subsection{Solver}
\label{sec:solver}
The \inline{tdgl.solve()} function implements the calculation described in Sections~\ref{sec:numerics} and \ref{appendix:pseudocode}. The required inputs to \inline{tdgl.solve()} are a \inline{tdgl.Device} instance for which the mesh has already been generated and an instance of \inline{tdgl.SolverOptions}. \inline{tdgl.SolverOptions} contains the total evolution time (in units of $\tau_0$), the parameters of the adaptive time step algorithm (see Section~\ref{sec:adaptive} and \ref{appendix:screening}), the units to use for magnetic fields and currents (e.g., \inline{"mT"} and \inline{"uA"}), and information about where and how often to save simulation results to disk.

Optional inputs to \inline{tdgl.solve()} include: \inline{applied_vector_potential}, a function that defines the applied magnetic vector potential\footnote{The parameter \inline{applied_vector_potential} can also be provided as a \inline{float} $B_z$, in which case it is interpreted to mean the vector potential associated with a uniform out-of-plane field with strength $B_z$ in units of the specified \inline{field_units}: $\mathbf{A}_\mathrm{applied}(\mathbf{r})=(1/2)B_z\hat{\mathbf{z}}\times\mathbf{r}$ in the symmetric gauge.} $\mathbf{A}_\mathrm{applied}(\mathbf{r})$; \inline{terminal_currents}, a constant or function that defines the applied bias currents $I_{\mathrm{ext},i}$ for any normal metal current terminals $i$; and \inline{disorder_epsilon}, a function that specifies the local critical temperature $\epsilon(\mathbf{r})$.
% The \inline{tdgl.sources} module provides functions that compute the vector potential for a uniform out-of-plane magnetic field and for a circular loop of current lying parallel to the plane of the film.
By default, the initial conditions of the system at time $t=0$ are $\psi(\mathbf{r}, 0)=1$ and $\mu(\mathbf{r}, 0)=0$ for all positions $\mathbf{r}$ not fixed by the boundary conditions (Section~\ref{sec:boundary}). Optionally, one can provide a ``seed solution'' generated by solving the same \inline{tdgl.Device}, potentially with a different applied vector potential or bias currents. The final state of the seed solution will then be used as the initial state for the new simulation.

\subsection{Post-processing}
\label{sec:solution}
The output of \inline{tdgl.solve()} is an instance of \inline{tdgl.Solution}, which encompasses all the information that defines the model and implements many methods to post-process and visualize the data generated during the simulation. Similarly to \inline{tdgl.Device}, \inline{tdgl.Solution} translates the results from the dimensionless units used for numerics into the physical units specified by the user. \inline{tdgl.Solution.tdgl_data} is a container for the raw data for a single solve step ($\psi$, $\mu$, $\mathbf{J}_s$, and $\mathbf{J}_n$ in dimensionless units). The raw data for a specific solve step \inline{N} can be loaded from disk by setting \inline{tdgl.Solution.solve_step = N} (\inline{N = -1} will load the final state). \inline{tdgl.Solution.dynamics} is an object that contains scalar data that are measured at each solve step $n$, namely, the time step $\Delta t^n$ and the electric potential $\mu$ and phase $\theta=\arg\psi$ measured at the device's ``probe points.'' Using \inline{tdgl.Solution.dynamics}, one can extract the dynamics of the voltage, $V_{ij}(t^n)=\mu(\mathbf{r}_{p,i}, t^n)-\mu(\mathbf{r}_{p,j}, t^n)$, and phase difference, $\Delta\theta_{ij}(t^n)=\theta(\mathbf{r}_{p,i}, t^n)-\theta(\mathbf{r}_{p,j}, t^n)$, between any two probe points $\mathbf{r}_{p,i}$ and $\mathbf{r}_{p,j}$. \inline{tdgl.Solution.dynamics.mean_voltage(i, j)} calculates the time-averaged voltage $\langle V_{ij}\rangle$ over a window of dimensionless time, which is useful for simulating current-voltage ($IV$) curves (see Section~\ref{sec:nanosquid}).

\inline{tdgl.Solution} has methods to evaluate the order parameter $\psi$ and the sheet current density $\mathbf{K}$ at any point within the device by interpolation. These methods can be used to calculate the total current $I_P$ crossing a given path $P$ in the device (via \inline{tdgl.Solution.current_through_path()}), or to evaluate the fluxoid $\Phi_C$ for a given closed curve $C$ in the device (via \inline{tdgl.Solution.polygon_fluxoid()} or \inline{tdgl.Solution.hole_fluxoid()}). The current through a path is given by
\begin{equation}
    \label{eq:I-through-path}
    I_P=\int_P\mathbf{K}(\mathbf{r})\cdot\hat{\mathbf{n}}(\mathbf{r})\,\mathrm{d}r,
\end{equation}
where $\hat{\mathbf{n}}(\mathbf{r})$ is a unit vector that is normal to the path $P$ at every position $\mathbf{r}.$

The fluxoid~\cite{Tinkham2004-ln} for a closed curve $C$ inside a superconducting film is given by
\begin{subequations}
\label{eq:fluxoid}
\begin{align}
    \Phi_C &= \Phi_C^\mathrm{flux} + \Phi_C^\mathrm{supercurrent}\\
    &= \oint_C\mathbf{A}(\mathbf{r})\cdot\mathrm{d}\mathbf{r}
        +\oint_C\mu_0\Lambda(\mathbf{r})\mathbf{K}_s(\mathbf{r})\cdot\mathrm{d}\mathbf{r}\label{eq:fluxoid-current}\\
    &=\frac{\Phi_0}{2\pi}\oint_C\nabla\theta(\mathbf{r})\cdot\mathrm{d}\mathbf{r}\label{eq:fluxoid-phase},
\end{align}
\end{subequations}
where $\mathbf{K}_s$ is the sheet supercurrent density, $\Lambda(\mathbf{r})=\Lambda_0/|\psi(\mathbf{r})|^2$ is the 
local effective magnetic penetration depth, $\Lambda_0=\lambda^2/d$ is the zero-field effective magnetic penetration depth, and $\theta(\mathbf{r})$ is the unwrapped phase of the order parameter. Because the phase of the order parameter (modulo $2\pi$) must be single-valued, $\Phi_C$ is quantized in units of $\Phi_0$. $\Phi_C/\Phi_0$ gives the signed number of vortices in the region enclosed by the curve $C$ (including those trapped in holes rather than in the film), or equivalently the number of $2\pi$ phase windings of the order parameter around the curve $C$.

Fluxoid quantization, $\Phi_C/\Phi_0\in\mathbb{Z}$, is automatically satisfied in GL theory, in contrast to London theory where it must be imposed as an external constraint. For this reason, fluxoid quantization provides a simple diagnostic to determine whether neglecting screening is a good approximation for a given model. If a solution generated under the assumption of negligible screening violates fluxoid quantization (where the fluxoid is evaluated using the supercurrent and vector potential, Eq.~\ref{eq:fluxoid-current}), the solution is not self-consistent and neglecting screening may not be a good approximation.

\inline{tdgl.Solution} also has methods to visualize a snapshot of the order parameter, electric potential, sheet current density, and current vorticity (the curl of the sheet current density) at a single solve step. To visualize dynamics in two dimensions, one can use \inline{tdgl.visualize}, which is a command line tool for interactive visualization and generating animations of time-dependent solutions. Solutions can be saved to disk using \inline{tdgl.Solution.to_hdf5()} and loaded from disk using \inline{tdgl.Solution.from_hdf5()}.

\begin{code-onecol}
\begin{minted}[fontsize=\small]{python}
options = tdgl.SolverOptions(
    solve_time=2000,
    field_units="mT",
)
fields = numpy.linspace(0.2, 1.9, 18)
solutions = []
for field in fields:
    solution = tdgl.solve(
        device,
        options,
        applied_vector_potential=field,
    )
    solutions.append(solution)

hole_fluxoids = []
for solution in solutions:
    solution.plot_order_parameter()
    hole_fluxoids.append(
        sum(solution.hole_fluxoid("hole"))
    )
\end{minted}
\captionof{listing}{Code used to model the superconducting ``py'' (Figure~\ref{fig:voronoi}) as a function of the uniform applied out-of-plane field, and visualize the order parameter as in Figure~\ref{fig:py-vs-field}. The name \inline{device} refers to the \inline{tdgl.Device} shown in Figure~\ref{fig:voronoi}. We also calculate via Eq.~\ref{eq:fluxoid-current} the fluxoid $\Phi_C/\Phi_0$ for a curve $C$ enclosing the hole (which is named \inline{"hole"}).}
\label{code:py-vortices}
\end{code-onecol}

\section{Examples}
\label{sec:examples}

\subsection{Vortices in irregularly shaped films}

\begin{figure}[h!]
    \centering
    \includegraphics[width=\linewidth]{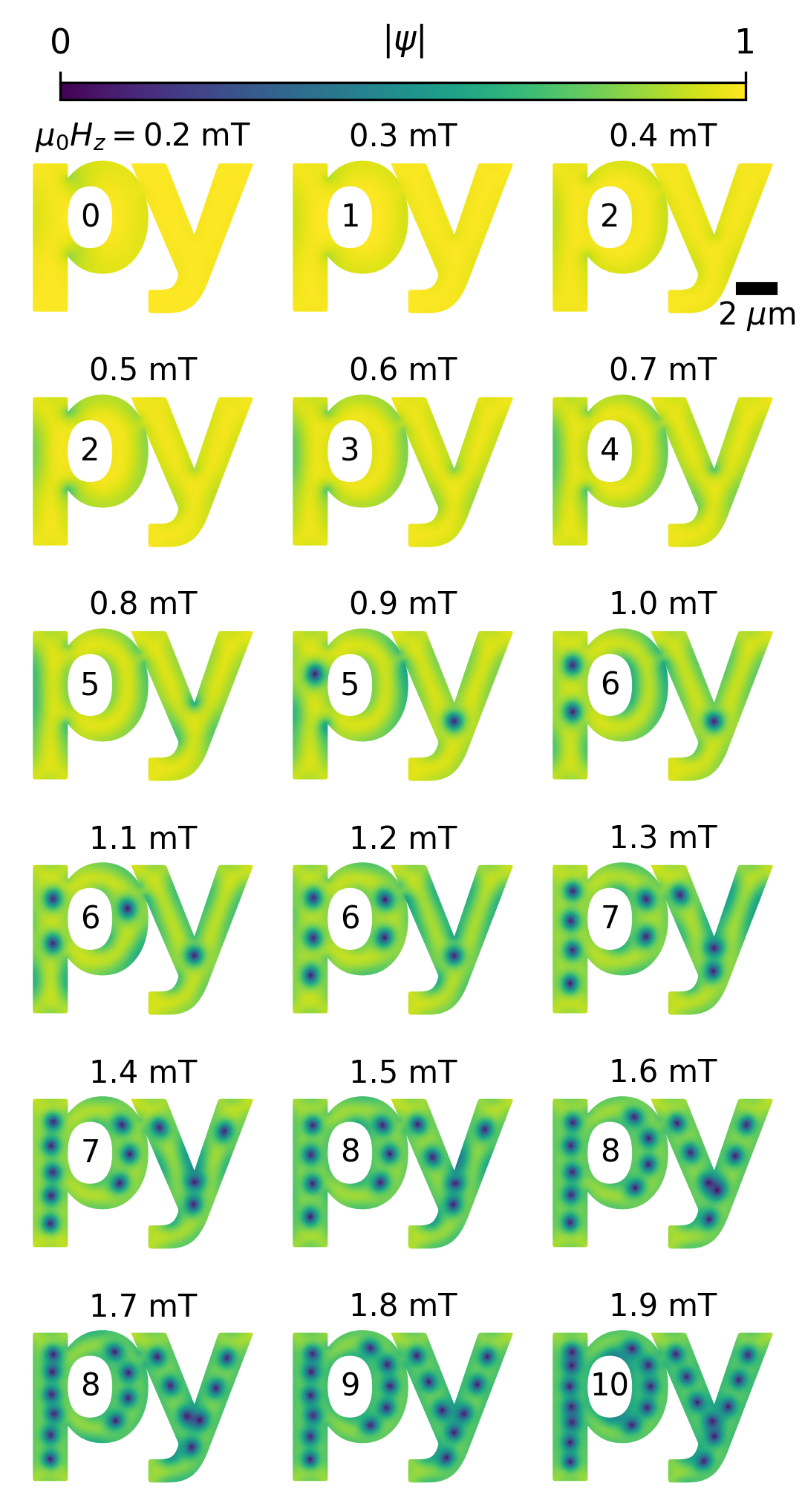}
    \caption{The magnitude of the order parameter, $|\psi|=\sqrt{n_s}$ as a function of uniform applied out-of-plane magnetic field $\mu_0H_z$ for a superconducting ``py'' with coherence length $\xi=0.4\,\um$, London penetration depth $\lambda=4\,\um$, film thickness $d=0.1\,\um$, and inelastic scattering parameter $\gamma=10$. The number of flux quanta trapped in the hole, found using \inline{tdgl.Solution.hole_fluxoid()}, is indicated by the number inside the ``p''. The total simulation time for each value of the applied field is $2,000\,\tau_0$.}
    \label{fig:py-vs-field}
\end{figure}

In the absence of applied transport currents, one can model the quasi-equilibrium behavior of a superconducting film using TDGL by allowing the system to evolve until the order parameter is roughly constant as a function of time.\footnote{``Quasi-equilibrium'' because we only allow the system to evolve for a finite amount of time.} To demonstrate, here we simulate the distribution of vortices in a mesoscopic superconductor patterned into the letters ``py'' (Figures~\ref{fig:voronoi} \& \ref{fig:py-vs-field}) and subject to a uniform out-of-plane applied magnetic field $\mu_0H_z$ (Code Block~\ref{code:py-vortices}).

For each value of the applied field, the film is initially in the Meissner state, with $|\psi|=1$ everywhere and zero $2\pi$ phase windings around the hole in the ``p''. A sheet supercurrent density $\mathbf{K}_s$ flows in the film, maintaining the zero-fluxoid condition $\Phi_C=0$ for all closed curves $C$ in the film, including curves that enclose the hole. $\mathbf{K}_s$ is inhomogeneous due to the shape of the film, and the superfluid density $n_s=|\psi|^2$ is locally suppressed in areas of the film where the supercurrent density approaches the critical current density, as can be seen in the top row of Figure~\ref{fig:py-vs-field}. Above some critical field, which depends on both the material parameters and the geometry of the film, it is energetically favorable for a phase slip to occur, allowing one flux quantum to enter the hole in the film. For the geometry and material parameters modeled here, as we increase the applied field, five vortices become trapped in the hole before it is energetically favorable for vortices to be trapped in the film.

\begin{figure*}[h!]
    \centering
    \includegraphics[width=\textwidth]{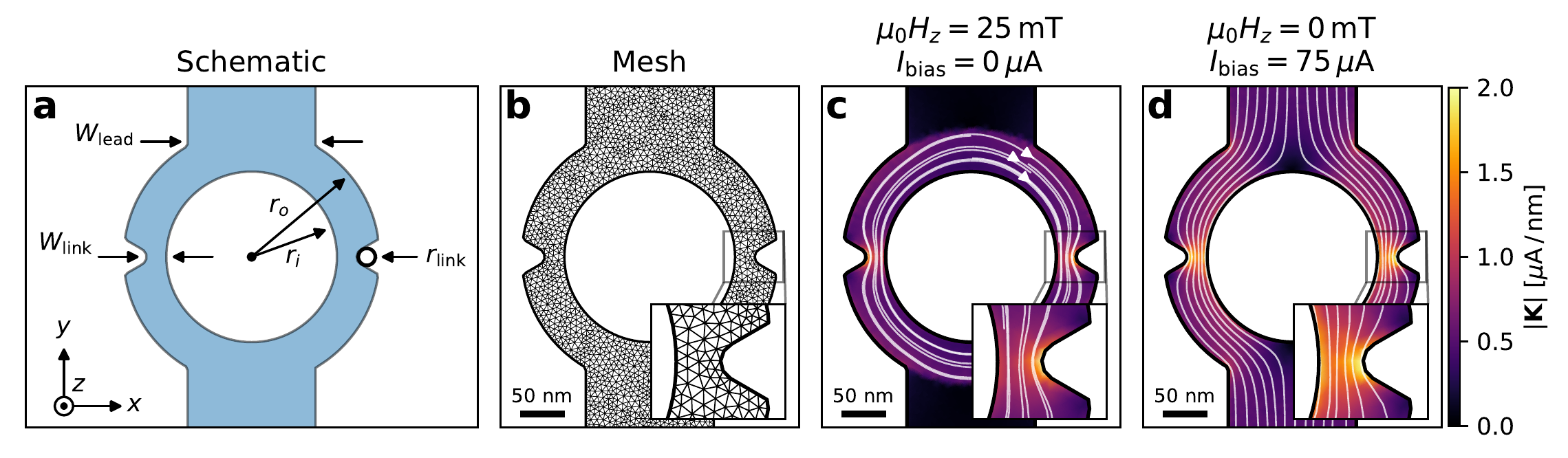}
    \caption{({\bf a}) Schematic of a weak link SQUID. The SQUID consists of an annulus with inner radius $r_i$ and outer radius $r_o$, interrupted by two rounded-triangular constrictions (weak links) with minimum width $W_\mathrm{link}$, radius of curvature $r_\mathrm{link}$, and included angle of $60^\circ$. A bias current $I_\mathrm{bias}$ can be applied through the two leads, each with width $W_\mathrm{lead}$. The leads extend beyond the top and bottom of the figure, and the normal metal contacts are not visible here. ({\bf b}) A finite volume mesh for the SQUID device, with $r_i=100\,\nm$, $r_o=W_\mathrm{lead}=150\,\nm$, $W_\mathrm{link}=25\,\mathrm{nm}$, and $r_\mathrm{link}=10\,\nm$. The mesh has a maximum edge length of $10\,\nm$. ({\bf c}) Simulated sheet current density $\mathbf{K}$ for zero bias current and an applied out-of-plane magnetic field $\mu_0H_z=25\,\mathrm{mT}$, assuming $\xi=50\,\nm$, $\lambda=200\,\nm$, $d=20\,\nm$, and $\gamma=10$. We assume that the leads (top and bottom) are angled at $\phi=75^\circ$ from the $x-y$ plane such that the component of the field perpendicular to the leads is $25\,\mathrm{mT}\times\cos75^\circ$. The flux threading the loop is approximately $\Phi_\mathrm{applied}=0.47\Phi_0$ and the magnitude of the clockwise-circulating screening current is $I_s=28\,\mu\mathrm{A}$. ({\bf d}) Simulated sheet current density for a bias current $I_\mathrm{bias}=75\,\mu\mathrm{A}$ flowing from top to bottom and zero applied magnetic field, assuming the same material parameter as ({\bf c}). The current is split evenly between the two sides of the SQUID. For both ({\bf c}) and ({\bf d}), the applied fields and currents are small enough that the device is fully superconducting, so $\mathbf{K}_n=0$ and $\mathbf{K}=\mathbf{K}_s$. The zero-field critical current for the SQUID is approximately $92\,\mu\mathrm{A}$ (Figure~\ref{fig:squid-dynamics}({\bf i})). Insets in ({\bf b}, {\bf c}, and {\bf d}) show an enlarged view of the right weak link.}
    \label{fig:squid-schematic}
\end{figure*}

Measurements of such irregularly shaped, micron-scale superconducting devices have proliferated in recent years due to the discovery of superconductivity in a variety of Van der Waals heterostructures, e.g. Refs~\cite{Cao2018-rx, Fatemi2018-az, Park2021-pk, Park2022-pu}, in which Fraunhofer-like current-voltage-field ($IVB$) characteristics are often used as supporting evidence for the presence of superconductivity. The ability to model vortex dynamics and IV characteristics in these mesoscopic devices, which often have many current terminals, may yield valuable insights into the nature of their superconductivity.

\subsection{nanoSQUID dynamics}
\label{sec:nanosquid}

The mechanism for dissipation within TDGL theory is the motion of vortices or the occurrence of phase slips within a superconducting film, driven by transport currents or a time-dependent applied magnetic field~\cite{Skocpol1974-sc, Watts-Tobin1981-mn, Ivlev1984-ct, Sivakov2003-ss}. In this section, we model dissipation caused by phase slips occurring at superconducting weak links. A weak link is a section of a superconducting film where the dimensions transverse to the direction of current flow are restricted, leading to an increase in the current density within the link and a corresponding decrease in the critical current of the link relative to the rest of the film~\footnote{A section of a superconducting film with a lower critical temperature than its surroundings can also act as a weak link.} (Figure~\ref{fig:squid-schematic}({\bf c},~{\bf d})). A weak link provides a nucleation site for phase slips of the order parameter and can therefore act as a Josephson junction~\cite{Vijay2009-is}, particularly when the size of the link is comparable to or smaller than the coherence length $\xi$. Such a weak link is known as a Dayem bridge~\cite{Dayem1967-ev}, nanobridge, or S-S'-S junction.

A nanoscale superconducting quantum interference device (nanoSQUID) consists of a small (typically $\sim$$100\,\nm$ -- $1\,\um$ diameter) superconducting loop interrupted by two weak links~\cite{Hasselbach2002-ii, Vasyukov2013-qh, Nulens2022-iq, Wyss2022-us}.\footnote{A nanoSQUID may also have more than two weak links~\cite{Anahory2014-xy, Uri2016-qg}.} As a simple model of a nanoSQUID, consider a flat annulus (loop) with effective area $A_\mathrm{eff}$ lying the $x-y$ plane, with leads angled out of the $x-y$ plane and two notches forming weak links on opposing sides of the loop (see Figure~\ref{fig:squid-schematic}({\bf a})). A uniform out-of-plane applied magnetic field $\mu_0\mathbf{H}=\mu_0H_z\hat{\mathbf{z}}$ will thread a flux $\Phi_\mathrm{applied}=\mu_0H_zA_\mathrm{eff}$ through the loop. In the absence of thermally-activated phase slips, if $|\Phi_\mathrm{applied}| < \Phi_0 / 2$, the loop will be in the Meissner (zero-fluxoid) state and a screening current $I_s$ will flow around the loop, the magnitude of which depends on the effective penetration depth $\Lambda$. We will define $I_s$ such that $I_s > 0$ indicates a counterclockwise circulating screening current.

\begin{figure}
    \centering
    \includegraphics[width=\linewidth]{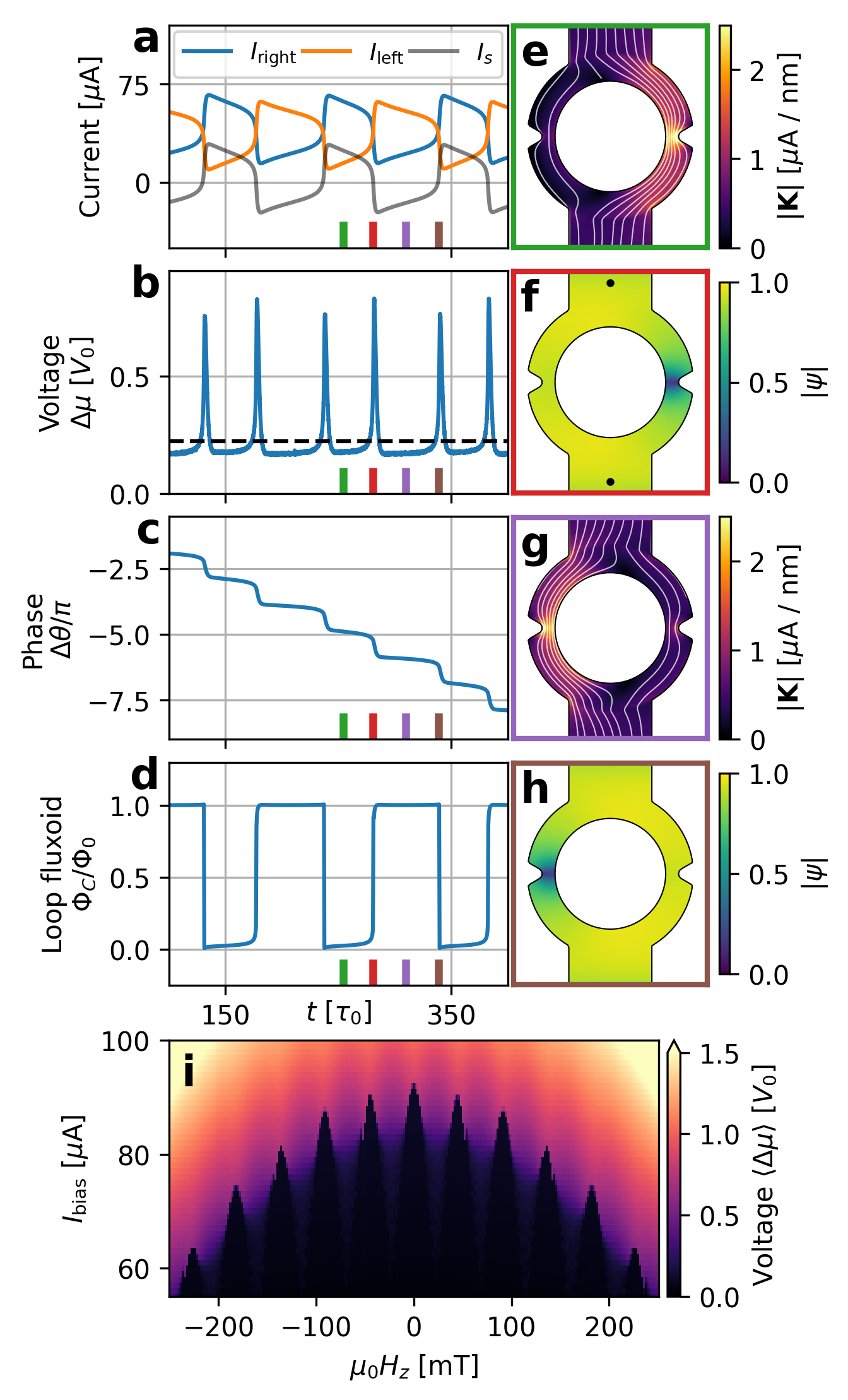}
    \caption{Simulated dynamics of the nanoSQUID shown in Figure~\ref{fig:squid-schematic}, with a uniform applied magnetic field $\mu_0H_z=25\,\mathrm{mT}$ and an applied bias current $I_\mathrm{bias}=75\,\mu\mathrm{A}$, flowing from top to bottom. ({\bf a})~The total current through the right (blue line) and left (orange line) legs of the SQUID, and the total circulating current (gray line, $I_s>0$ is clockwise). ({\bf b})~The voltage $V=\Delta\mu$ across the SQUID, measured between the two probe points indicated by black dots in ({\bf f}). The voltage is characterized by short spikes, each corresponding to a phase slip across one of the weak links. The phase slips alternate between the two weak links. The mean voltage is indicated by the dashed black line. ({\bf c})~The unwrapped phase difference between the two probe points. The phase difference advances by $2\pi$ each time there is a phase slip across both junctions, corresponding to one $\Phi_0$ being transferred across the device. ({\bf d})~The loop fluxoid, i.e., the number of flux quanta trapped in the hole of the SQUID. ({\bf e}, {\bf g})~The sheet current density at the times indicated by the green and purple bars in ({\bf a}-{\bf d}), corresponding to a loop fluxoid of $0\,\Phi_0$ and $1\,\Phi_0$ respectively. ({\bf f}, {\bf h})~The magnitude of the order parameter $|\psi|=\sqrt{n_s}$ at the times indicated by the red and brown bars in ({\bf a}-{\bf d}), corresponding to a phase slip across the right and left weak link respectively. ({\bf i})~Simulated current-voltage-flux ($IVB$) characteristic of the nanoSQUID. The color scale indicates the time-averaged voltage $\langle\Delta\mu\rangle$ across the device as a function of bias current and applied magnetic field, showing clear SQUID-like oscillations in the critical current, with periodicity consistent with an effective loop radius of $r_\mathrm{eff}=120\,\mathrm{nm}$. The color scale is truncated at $1.5\,V_0$.}
    \label{fig:squid-dynamics}
\end{figure}

Suppose $0 < \Phi_\mathrm{applied} < \Phi_0 / 2$ so that $I_s$ flows clockwise around the loop (Figure~\ref{fig:squid-schematic}({\bf c})). Due to this screening current, applying a constant bias current $I_\mathrm{bias}$ flowing from the top lead to the bottom lead will result in a larger total current in the right weak link ($I_\mathrm{right}=I_\mathrm{bias} / 2 + I_s > I_\mathrm{bias} / 2$) than in the left weak link ($I_\mathrm{left}=I_\mathrm{bias} / 2 - I_s < I_\mathrm{bias} / 2$) (Figure~\ref{fig:squid-dynamics}({\bf e})). If $I_\mathrm{right}$ exceeds the critical current of the right link, a phase slip will occur across the link, allowing $\Phi_0$ to enter the hole in the loop and generating a short voltage pulse across the device (Figure~\ref{fig:squid-dynamics}({\bf b}, {\bf c}, {\bf d}, {\bf f})). Now that one flux quantum is trapped inside the hole, the screening current will redistribute to maintain the fluxoid state $\Phi_C/\Phi_0=1$, causing $I_s$ to change sign so that $I_\mathrm{right}=I_\mathrm{bias} / 2 + I_s < I_\mathrm{bias} / 2$ and $I_\mathrm{left}=I_\mathrm{bias} / 2 - I_s > I_\mathrm{bias} / 2$ (Figure~\ref{fig:squid-dynamics}({\bf g})). Now, if $I_\mathrm{left}$ exceeds the critical current of the left link, a phase slip will occur across the left link, allowing one $\Phi_0$ to exit the hole in the loop, generating a short voltage pulse across the device, and restoring the $\Phi_C/\Phi_0=0$ fluxoid state (Figure~\ref{fig:squid-dynamics}({\bf b}, {\bf c}, {\bf d}, {\bf h})). In the absence of self-heating, this cycle will repeat indefinitely, and a DC measurement of the voltage across the SQUID will measure the time-average of the series of voltage pulses (dashed line in Figure~\ref{fig:squid-dynamics}({\bf b})). The larger the bias current, the shorter the time interval between phase slips and, therefore, the larger the time-averaged voltage. To convert the time and voltage to physical units, one must specify the normal state conductivity $\sigma$ of the film (see Section~\ref{sec:units}). For example, assuming a normal state conductivity of $\sigma=(0.1\,\mu\Omega\cdot\mathrm{cm})^{-1}=10^3\,\mathrm{S}/\um$, we have $\tau_0 \approx 50\,\mathrm{ps}$ and $V_0\approx 26\,\mathrm{mV}$.

Figure~\ref{fig:squid-dynamics}({\bf i}) shows the simulated current-voltage-field ($IVB$) characteristic of the nanoSQUID. The color scale indicates the time-averaged voltage $\langle\Delta\mu\rangle$ across the device as a function of bias current $I_\mathrm{bias}$ and applied magnetic field $\mu_0H_z$, showing clear SQUID-like oscillations of the critical current as a function of applied field. The periodicity of the oscillations is consistent with an effective loop area of $A_\mathrm{eff}=\pi r_\mathrm{eff}^2$ with $r_\mathrm{eff}=120\,\mathrm{nm}$, in good agreement with the dimensions of the device. In addition to the SQUID oscillations, there is an overall decrease in the critical current with increasing magnetic field strength, as seen in experimental $IVB$ characteristics of nanoSQUID devices, for instance in Refs.~\cite{Vasyukov2013-qh, Wyss2022-us}. Simulations like this can be used to optimize the design of nanoSQUIDs and similar devices such as superconducting nanowire single photon detectors (SNSPDs)~\cite{Berdiyorov2012-rn, Jonsson2022-mb}.

\section{Conclusion}
\label{sec:conclusion}
\pyTDGL is a user-friendly, portable, and thoroughly documented implementation of a widely used generalized time-dependent Ginzburg-Landau method~\cite{Kramer1978-kb,Watts-Tobin1981-mn}. The model is applicable to dirty two-dimensional superconductors in the limit of weak screening and for temperatures very close to the critical temperature. Under these assumptions, \pyTDGL can simulate the magnetic response and dynamics of superconducting films of any geometry, including films with holes and films with multiple current terminals. The package includes methods for analyzing and visualizing simulation results as a function of both space and time. These capabilities make \pyTDGL a valuable addition to the computational toolbox for modeling superconductors and superconducting devices, which includes open-source~\cite{Bishop-Van_Horn2022-sy}, closed-source~\cite{Khapaev2001-pw}, and commercial~\cite{Fourie2011-wl} London-Maxwell solvers, 2D~\cite{Alstrom2011-bc, Jonsson2022-mb} and 3D~\cite{Oripov2020-dq, Peng2017-zt, Sadovskyy2015-ha, Winiecki2002-ka} TDGL solvers, and solvers for alternative theories such as the quasiclassical theory of superconductivity~\cite{Holmvall2022-ps, Seja2022-aa}.

We hope that \pyTDGL will accelerate the design and optimization of superconducting devices and enhance the transferability of research results by lowering the entry barrier for realistic modeling of both transport and scanning-probe measurements. \pyTDGL may also be a valuable tool for teaching and building intuition about dynamics and dissipation in 2D superconductors. Possible extensions to \pyTDGL include coupling to equations of motion for other degrees of freedom, such as the heat transport equation to model self-heating induced by vortex motion~\cite{Gurevich1987-sv, Berdiyorov2012-rn, Zotova2012-nc, Jelic2016-ww, Jing2018-qc}. Future work may also include optimizations to the adaptive time step (Section~\ref{sec:adaptive}) and screening (\ref{appendix:screening}) algorithms, and the addition of stochastic terms to model thermal or electrical noise~\cite{Sadovskyy2015-ha}.

\section{Acknowledgements}
\label{section:acknowledgements}
This work is supported by the Department of Energy, Office of Science, Basic Energy Sciences, Materials Sciences and Engineering Division, under Contract DE-AC02-76SF00515. Some of the computing for this project was performed on the Sherlock cluster. We would like to thank Stanford University and the Stanford Research Computing Center for providing computational resources and support that contributed to these research results. We would also like to thank the authors of Ref.~\cite{Jonsson2022-mb} for sharing their code in a public GitHub repository, \href{https://github.com/afsa/super-detector-py}{https://github.com/afsa/super-detector-py}, which served as inspiration and a starting point for \pyTDGL.

% \newpage
\appendix

\section{Euler method}
\label{appendix:euler}

Below we derive the quadratic equation for $\left|\psi_i^{n+1}\right|^2$,
Eq.~\ref{eq:quad-2}, starting from Eq.~\ref{quad-1} (reproduced below as Eq.~\ref{eq:quad-1-1}. This calculation is the basis for the Euler method used in \pyTDGL.

\begin{equation}
    \label{eq:quad-1-1}
    \psi_i^{n+1} = w_i^{n} - z_i^{n}\left|\psi_i^{n+1}\right|^2
\end{equation}
\begin{equation}
    \label{quad-full}
    \begin{split}
        \left|\psi_i^{n+1}\right|^2 =& \left(\psi_i^{n+1}\right)^*\left(\psi_i^{n+1}\right)\\
        =& \left(w_i^{n}-z_i^{n}\left|\psi_i^{n+1}\right|^2\right)^*\left(w_i^{n}-z_i^{n}\left|\psi_i^{n+1}\right|^2\right)\\
        =& \left|w_i^{n}\right|^2 \\
        & - {w_i^{n}}^*z_i^{n}\left|\psi_i^{n+1}\right|^2\\
        & - w_i^{n}{z_i^{n}}^*\left|\psi_i^{n+1}\right|^2 \\
        & + \left|z_i^{n}\right|^2\left|\psi_i^{n+1}\right|^4
        % \left|\psi_i^{n+1}\right|^2\left(1 + {w_i^{n}}^*z_i^{n} + w_i^{n}{z_i^{n}}^*\right)
        % =&\left|w_i^{n}\right|^2 + \left|z_i^{n}\right|^2\left|\psi_i^{n+1}\right|^4
        % {w_i^{n}}^*z_i^{n} + w_i^{n}{z_i^{n}}^* =& 2\left(\mathrm{Re}\{w_i^{n}\}\mathrm{Re}\{z_i^{n}\}+\mathrm{Im}\{w_i^{n}\}\mathrm{Im}\{z_i^{n}\}\right)\\
        % =& 2c_i^{n}\\
        % 0 =& \left|z_i^{n}\right|^2\left|\psi_i^{n+1}\right|^4 - (2c_i^{n} + 1)\left|\psi_i^{n+1}\right|^2 + \left|w_i^{n}\right|^2
    \end{split}
\end{equation}
\begin{equation}
        \left|\psi_i^{n+1}\right|^2\left(1 + {w_i^{n}}^*z_i^{n} + w_i^{n}{z_i^{n}}^*\right)
        =\left|w_i^{n}\right|^2 + \left|z_i^{n}\right|^2\left|\psi_i^{n+1}\right|^4
\end{equation}
Using the identity
\begin{equation}
    \begin{split}
        {w_i^{n}}^*z_i^{n} + w_i^{n}{z_i^{n}}^* =& 2\left(\mathrm{Re}\{w_i^{n}\}\mathrm{Re}\{z_i^{n}\}+\mathrm{Im}\{w_i^{n}\}\mathrm{Im}\{z_i^{n}\}\right)\\
        =& 2c_i^{n},
    \end{split}
\end{equation}
we arrive at a quadratic equation in $\left|\psi_i^{n+1}\right|^2$:
\begin{equation}
    \left|z_i^{n}\right|^2\left|\psi_i^{n+1}\right|^4 - (2c_i^{n} + 1)\left|\psi_i^{n+1}\right|^2 + \left|w_i^{n}\right|^2=0,
\end{equation}
which we solve using Eq.~\ref{eq:citardauq}.

\section{Robustness of the adaptive algorithm}
\label{appendix:adpative}

\begin{figure}[h]
    \centering
    \includegraphics[width=\linewidth]{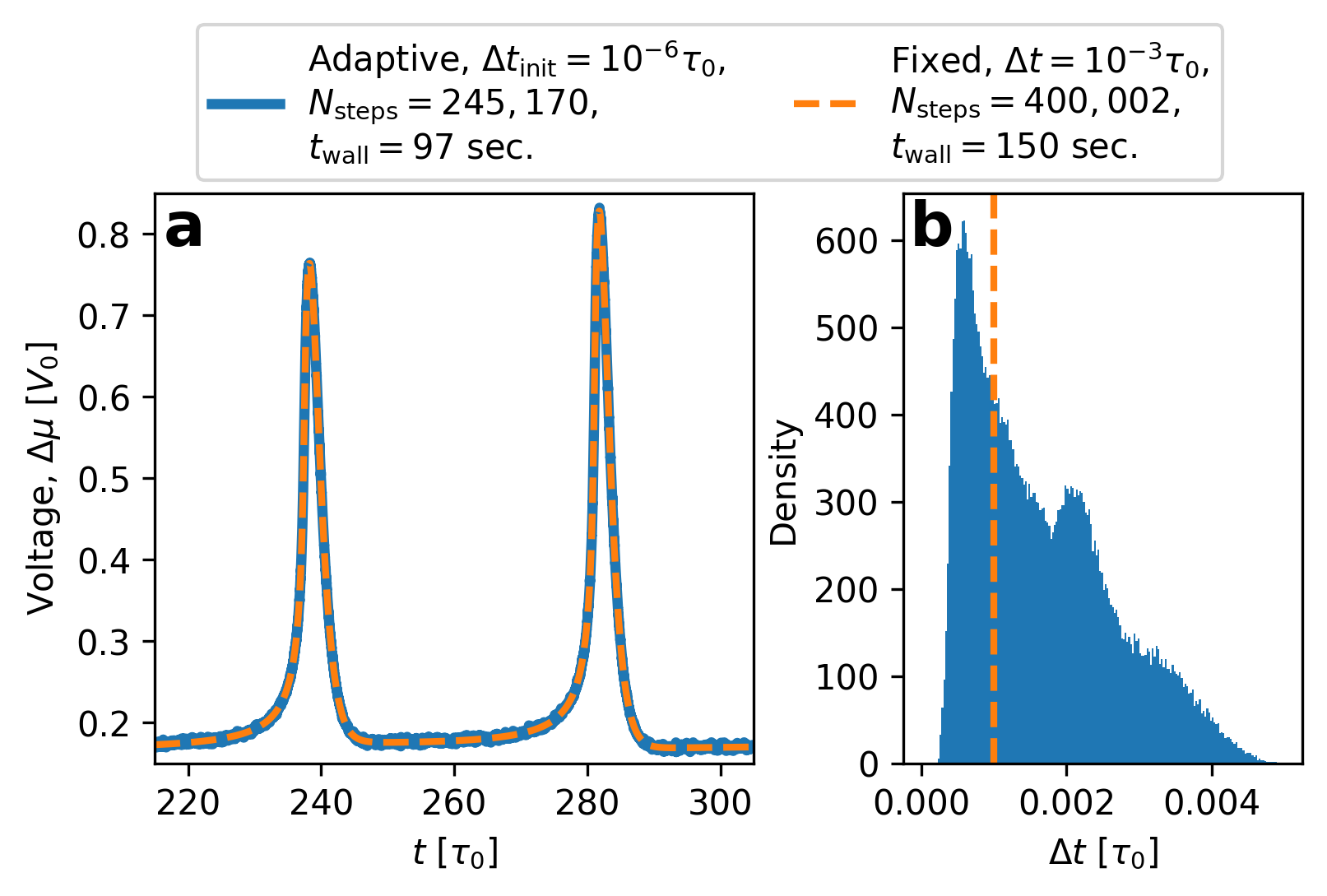}
    \caption{Insensitivity of simulated dynamics to the parameters of the adaptive time step algorithm. ({\bf a}) Voltage as a function of time for the nanoSQUID modeled in  Section~\ref{sec:nanosquid}. The solid blue curve shows the results using an adaptive time step with $\Delta t_\mathrm{init}=10^{-6}\tau_0$, $\Delta t_\mathrm{max}=10^{-1}\tau_0$, and $N_\mathrm{window}=10$ (this is the same data as Figure~\ref{fig:squid-dynamics}({\bf b})). The dashed orange curve shows the results for a fixed time step of $\Delta t=10^{-3}\tau_0$, which is the largest fixed time step for which the simulation would converge. ({\bf b}) A normalized histogram of the time steps $\Delta t^n$ for the adaptive case, with the fixed time step $\Delta t=10^{-3}\tau_0$ indicated by the vertical dashed orange line. The number of solve steps $N_\mathrm{steps}$ and the wall clock time $t_\mathrm{wall}$ required to simulate a total evolution time of $400\tau_0$ in each case is given in the figure legend.}
    \label{fig:adaptive}
\end{figure}

The simulated dynamics in a \pyTDGL model are in general quite insensitive to the exact parameters chosen for the adaptive time step algorithm (Section~\ref{sec:adaptive}). Figure~\ref{fig:adaptive} shows the dynamics of the nanoSQUID examined in Section~\ref{sec:nanosquid}, simulated using both an adaptive time step (as in Figure~\ref{fig:squid-dynamics}) and a fixed time step for a total evolution time of $400\tau_0$. The resulting voltage is nearly indistinguishable between the two cases, and the adaptive case requires roughly 39\% fewer solve steps and 35\% less wall clock time to complete the simulation. The size of the speedup from using an adaptive time step of course depends on the details of the model.

\section{Screening}
\label{appendix:screening}

As mentioned in Section~\ref{section:model-tdgl}, by default \texttt{pyTDGL} assumes that screening is negligible, i.e., that the total vector potential in the film is time-independent and equal to the applied vector potential: $\mathbf{A}(\mathbf{r}, t)=\mathbf{A}_\mathrm{applied}(\mathbf{r})$. This assumption is easily justified when the effective penetration depth of the film $\Lambda=\lambda^2/d$ is much larger than both the film thickness $d$ and the lateral size of the film~\cite{Tinkham2004-ln}. Screening can optionally be included by evaluating the vector potential induced by currents flowing in the film. The vector potential in a 2D film induced by a sheet current density $\mathbf{K}$ flowing in the film is given by
\begin{equation}
    \label{eq:A-induced}
    \mathbf{A}_\mathrm{induced}(\mathbf{r}, t) = \frac{\mu_0}{4\pi}\int_\mathrm{film}\frac{\mathbf{K}(\mathbf{r}', t)}{|\mathbf{r}-\mathbf{r}'|}\,\mathrm{d}^2\mathbf{r}'.
\end{equation}
Taking the induced vector potential into account, the total vector potential in the film is
\begin{equation}
    \label{eq:A-total}
    \mathbf{A}(\mathbf{r}, t)=\mathbf{A}_\mathrm{applied}(\mathbf{r})+\mathbf{A}_\mathrm{induced}(\mathbf{r}, t).
\end{equation}
Because $\mathbf{A} =\mathbf{A}_\mathrm{applied}+\mathbf{A}_\mathrm{induced}$ enters into the covariant gradient and Laplacian of $\psi$ (Eqs.~\ref{grad-psi} and \ref{lapacian-psi}), which in turn determine the current density $\mathbf{J}=\mathbf{K}/d$, which determines $\mathbf{A}_\mathrm{induced}$, Eq.~\ref{eq:A-induced} must be solved self-consistently at each time step $t^n$. The strategy for updating the induced vector potential to converge to a self-consistent value is based on Polyak's ``heavy ball'' method~\cite{Polyak1964-gb, Holmvall2022-ps}:
\begin{subequations}
    \label{eq:polyak}
    \begin{align}
        \mathbf{A}^{n,s}_{\mathrm{induced},ij} &= \frac{\mu_0}{4\pi}\sum_{\text{sites } \ell}\frac{\mathbf{K}^{n,s}_\ell}{|\mathbf{r}_{ij}-\mathbf{r}_\ell|}a_\ell\label{eq:polyak-A}\\
        \mathbf{d}^{n,s}_{ij} &= \mathbf{A}^{n,s}_{\mathrm{induced},ij} - \mathbf{A}^{n,s-1}_{\mathrm{induced},ij}\\
        \mathbf{v}^{n,s+1} &= (1-\beta)\mathbf{v}^{n,s} + \alpha\mathbf{d}^{n,s}_{ij}\label{eq:polyak-velocity}\\
        \mathbf{A}^{n,s+1}_{\mathrm{induced},ij} &= \mathbf{A}^{n,s}_{\mathrm{induced},ij} + \mathbf{v}^{n,s+1}_{ij}
    \end{align}
\end{subequations}
The integer index $s$ counts the number of iterations performed in the self-consistent calculation. The parameters $\alpha\in(0,\infty)$ and $\beta\in(0,1)$ in Eq.~\ref{eq:polyak-velocity} can be set by the user, and the initial conditions for Eq.~\ref{eq:polyak} are $\mathbf{A}^{n,0}_{\mathrm{induced},ij} = \mathbf{A}^{n-1}_{\mathrm{induced},ij}$ and $\mathbf{v}^{n,0}_{ij} = \mathbf{0}$. The iterative application of Eq.~\ref{eq:polyak} terminates when the relative change in the induced vector potential between iterations falls below a user-defined tolerance.

In Eq.~\ref{eq:polyak-A}, we evaluate the sheet current density $\mathbf{K}^n_\ell=\mathbf{K}(\mathbf{r}_\ell,t^n)$ on the mesh sites $\mathbf{r}_\ell$, and the vector potential on the mesh edges $\mathbf{r}_{ij}$, so the denominator $|\mathbf{r}_{ij}-\mathbf{r}_\ell|$ is strictly greater than zero and Eq.~\ref{eq:polyak-A} is well-defined. Eq.~\ref{eq:polyak-A} involves the pairwise distances between all edges and all sites in the mesh, so, in contrast to the sparse finite volume calculation, it requires a dense matrix representation. This means that including screening significantly increases both the memory footprint and the number of floating point operations required for a TDGL simulation. To accelerate this part of the calculation, Eq.~\ref{eq:polyak-A} is automatically evaluated on a graphics processing unit (GPU) if one is available. Although including screening does introduce some time-dependence to the total vector potential in the film (Eq.~\ref{eq:A-total}), we assume that $\partial\mathbf{A}/\partial t$ remains small enough that the electric field in the film is $\mathbf{E}=-\nabla\mu - \partial\mathbf{A}/\partial t \approx -\nabla\mu$. The screening calculation (Eq.~\ref{eq:polyak}) tends to converge slowly, and can fail to converge for models in which the effective magnetic penetration depth $\Lambda=\lambda^2/d$ is comparable to or smaller than the film size. For these reasons, it may be expedient to use other tools when screening cannot be neglected, for example, the tools presented in Refs.~\cite{Bishop-Van_Horn2022-sy, Holmvall2022-ps}. Pseudocode for the main solver when screening is included is given in Algorithm~\ref{alg:adaptive-screening} in \ref{appendix:pseudocode}.

\section{Pseudocode}
\label{appendix:pseudocode}

Here, we provide pseudocode for critical components of the \pyTDGL package.

\begin{algorithm}
    \caption{Adaptive Euler update. The parameters $M_\mathrm{adaptive}$ and $N_\mathrm{retries}^\mathrm{max}$ can be set by the user.}
    \label{alg:adaptive-euler-step}
    \KwData{$\psi_i^n$, $\Delta t_\star$, $M_\mathrm{adaptive}$, $N_\mathrm{retries}^\mathrm{max}$}
    \KwResult{$\psi_i^{n+1}$, $\Delta t^n$}
    $\Delta t^n \gets \Delta t_\star$\;
    Calculate $z_i^n$, $w_i^n$, $\left|\psi_i^{n+1}\right|^2$ given $\Delta t^n$ (Eqs.~\ref{eq:z}, \ref{eq:w}, \ref{eq:quad-root})\;
    \If{adaptive}{
        $N_\mathrm{retries} \gets 0$\;
        \While{$\left|\psi_i^{n+1}\right|^2$ is complex for any site $i$}{
            \If{$N_\mathrm{retries} > N_\mathrm{retries}^\mathrm{max}$}{
                Failed to converge - raise an error.
            }
            $\Delta t^n \gets \Delta t^n \times M_\mathrm{adaptive}$\;
            Calculate $z_i^n$, $w_i^n$, $\left|\psi_i^{n+1}\right|^2$ given $\Delta t^n$ (Eqs.~\ref{eq:z}, \ref{eq:w}, \ref{eq:quad-root})\;
            $N_\mathrm{retries} \gets N_\mathrm{retries} + 1$\;
        }
    }
    $\psi_i^{n+1} \gets w_i^n - z_i^n \left|\psi_i^{n+1}\right|^2$ (Eq.~\ref{eq:psi-sol})\;
\end{algorithm}

\begin{algorithm}
    \caption{A single solve step, in which we solve for the state of the system at time $t^{n+1}$ given the state of the system at time $t^n$, with no screening.}
    \label{alg:adaptive}
    \KwData{$n$, $t^n$, $\Delta t_\star$, $\psi_i^{n}$, $\mu_i^{n}$}
    \KwResult{$t^{n+1}$, $\Delta t^{n}$, $\psi_i^{n+1}$, $\mu_i^{n+1}$, $J_{s,ij}^{n+1}$, $J_{n,ij}^{n+1}$, $\Delta t_\star$}
    Evaluate current density $J^{n+1}_{\mathrm{ext},\,k}$ for terminals $k$ (Eq.~\ref{eq:bc-current})\;
    Update boundary conditions for $\mu_i^{n+1}$ (Eq.~\ref{eq:bc-normal-mu})\;
    Calculate $\psi_i^{n+1}$ and $\Delta t^n$ via Algorithm~\ref{alg:adaptive-euler-step}\;
    Calculate the supercurrent density $J_{s,ij}^{n+1}$ (Eq.~\ref{eq:supercurrent})\;
    Solve for $\mu_i^{n+1}$ via sparse LU factorization (Eq.~\ref{poisson-num})\;
    Evaluate normal current density $J_{n,ij}^{n+1}$ via $\mathbf{J}_n=-\nabla\mu$\;
    \If{adaptive}{
        Select new tentative time step $\Delta t_\star$ given $\Delta t^n$ (Eq.~\ref{eq:dt-tentative})\;
    }
    $t^{n+1} \gets t^{n} + \Delta t^{n}$\;
    $n \gets n + 1$\;
\end{algorithm}

\begin{algorithm}[h!]
    \caption{A single solve step, with screening. The parameters $A_\mathrm{tol}$ and $N_\mathrm{screening}^\mathrm{max}$ can be set by the user.}
    \label{alg:adaptive-screening}
    \KwData{$n$, $t^n$, $\Delta t_\star$, $\psi_i^{n}$, $\mu_i^{n}$, $\mathbf{A}^n_{\mathrm{induced}}$, $A_\mathrm{tol}$, $N^\mathrm{max}_\mathrm{screening}$}
    \KwResult{$t^{n+1}$, $\Delta t^{n}$, $\psi_i^{n+1}$, $\mu_i^{n+1}$, $J_{s,ij}^{n+1}$, $J_{n,ij}^{n+1}$, $\mathbf{A}^{n+1}_{\mathrm{induced}}$, $\Delta t_\star$}
    Evaluate current density $J^{n+1}_{\mathrm{ext},\,k}$ for terminals $k$ (Eq.~\ref{eq:bc-current})\;
    Update boundary conditions for $\mu_i^{n+1}$ (Eq.~\ref{eq:bc-normal-mu})\;
    $s \gets 0$, screening iteration index\;
    $\mathbf{A}^{n+1,s}_\mathrm{induced} \gets \mathbf{A}^{n}_\mathrm{induced}$, initialize induced vector potential based on solution from previous time step\;
    $\delta A_\mathrm{induced} \gets \infty$, relative error in induced vector potential\;
    \While{$\delta A_\mathrm{induced} > A_\mathrm{tol}$}{
        \If{$s > N_\mathrm{screening}^\mathrm{max}$}{
            Failed to converge - raise an error.
        }
        \If{$s == 0$}{
            $\Delta t^n \gets \Delta t_\star$, initial guess for new time step\;
        }
        Update link variables in $(\nabla-i\mathbf{A})$ and $(\nabla -i\mathbf{A})^2$ given $\mathbf{A}_\mathrm{induced}^{n+1,s}$ (Eqs.~\ref{grad-psi} and \ref{lapacian-psi})\;
        Calculate $\psi_i^{n+1}$ and $\Delta t^n$ via Algorithm~\ref{alg:adaptive-euler-step}\;
        Calculate the supercurrent density $J_{s,ij}^{n+1}$ (Eq.~\ref{eq:supercurrent})\;
        Solve for $\mu_i^{n+1}$ via sparse LU factorization (Eq.~\ref{poisson-num})\;
        Evaluate normal current density $J_{n,ij}^{n+1}$ via $\mathbf{J}_n=-\nabla\mu$\;
        Evaluate $\mathbf{K}_i^{n+1}=d(\mathbf{J}_{s,i}^{n+1}+\mathbf{J}_{n,i}^{n+1})$ at the mesh sites $i$\;
        Update induced vector potential $\mathbf{A}^{n+1,s}_\mathrm{induced}$ (Eq.~\ref{eq:polyak})\;
        \If{$s > 1$}{
            $\delta A_\mathrm{induced} \gets \max_\mathrm{edges}\left(\left|\mathbf{A}^{n+1,s}_\mathrm{induced}-\mathbf{A}^{n+1,s-1}_\mathrm{induced}\right|/\left|\mathbf{A}^{n+1,s}_\mathrm{induced}\right|\right)$
        }
        $s \gets s + 1$\;
    }
    $\mathbf{A}^{n+1}_\mathrm{induced} \gets \mathbf{A}^{n+1,s}_\mathrm{induced}$, self-consistent value of the induced vector potential\;
    \If{adaptive}{
        Select new tentative time step $\Delta t_\star$ (Eq.~\ref{eq:dt-tentative})\;
    }
    $t^{n+1} \gets t^{n} + \Delta t^{n}$\;
    $n \gets n + 1$\;
\end{algorithm}

%% References

% \newpage
\bibliographystyle{elsarticle-num}
\bibliography{references}

\begin{thebibliography}{10}
\expandafter\ifx\csname url\endcsname\relax
  \def\url#1{\texttt{#1}}\fi
\expandafter\ifx\csname urlprefix\endcsname\relax\def\urlprefix{URL }\fi
\expandafter\ifx\csname href\endcsname\relax
  \def\href#1#2{#2} \def\path#1{#1}\fi

\bibitem{Ginzburg2008-qb}
V.~L. Ginzburg, L.~D. Landau,
  \href{https://link.springer.com/chapter/10.1007/978-3-540-68008-6_4}{On the
  theory of superconductivity}, in: On Superconductivity and Superfluidity,
  Springer Berlin Heidelberg, Berlin, Heidelberg, 2008, pp. 113--137.
\newline\urlprefix\url{https://link.springer.com/chapter/10.1007/978-3-540-68008-6_4}

\bibitem{Bardeen1957-af}
J.~Bardeen, L.~N. Cooper, J.~R. Schrieffer,
  \href{https://link.aps.org/doi/10.1103/PhysRev.108.1175}{Theory of
  superconductivity}, Phys. Rev. 108~(5) (1957) 1175--1204.
\newline\urlprefix\url{https://link.aps.org/doi/10.1103/PhysRev.108.1175}

\bibitem{Bardeen1957-og}
J.~Bardeen, L.~N. Cooper, J.~R. Schrieffer,
  \href{https://link.aps.org/doi/10.1103/PhysRev.106.162}{Microscopic theory of
  superconductivity}, Phys. Rev. 106~(1) (1957) 162--164.
\newline\urlprefix\url{https://link.aps.org/doi/10.1103/PhysRev.106.162}

\bibitem{Gorkov1959-iv}
L.~P. Gor'kov, \href{https://www.osti.gov/biblio/7264935}{Microscopic
  derivation of the {Ginzburg--Landau} equations in the theory of
  superconductivity}, Sov. Phys. - JETP (Engl. Transl.); (United States) 9:6
  (Jan. 1959).
\newline\urlprefix\url{https://www.osti.gov/biblio/7264935}

\bibitem{Tinkham2004-ln}
M.~Tinkham,
  \href{https://play.google.com/store/books/details?id=JOQoAwAAQBAJ}{Introduction
  to Superconductivity: Second Edition}, Courier Corporation, 2004.
\newline\urlprefix\url{https://play.google.com/store/books/details?id=JOQoAwAAQBAJ}

\bibitem{Gorkov1996-do}
L.~P. Gor'kov, G.~M. {\'E}liashberg,
  \href{https://doi.org/10.1142/9789814317344_0003}{{Generalization of the
  Ginzurg-Landau Equations for Non-stationary Problems in the Case of Alloys
  With Paramagnetic Impurities}}, in: 30 Years of the Landau Institute:
  Selected Papers, Vol.~11 of World Scientific Series in 20th Century Physics,
  WORLD SCIENTIFIC, 1996, pp. 16--22.
\newline\urlprefix\url{https://doi.org/10.1142/9789814317344_0003}

\bibitem{Schmid1966-bh}
A.~Schmid, \href{https://doi.org/10.1007/BF02422669}{A time dependent
  {Ginzburg-Landau} equation and its application to the problem of resistivity
  in the mixed state}, Physik der kondensierten Materie 5~(4) (1966) 302--317.
\newline\urlprefix\url{https://doi.org/10.1007/BF02422669}

\bibitem{Kramer1978-kb}
L.~Kramer, R.~J. Watts-Tobin,
  \href{https://link.aps.org/doi/10.1103/PhysRevLett.40.1041}{Theory of
  dissipative {Current-Carrying} states in superconducting filaments}, Phys.
  Rev. Lett. 40~(15) (1978) 1041--1044.
\newline\urlprefix\url{https://link.aps.org/doi/10.1103/PhysRevLett.40.1041}

\bibitem{Watts-Tobin1981-mn}
R.~J. Watts-Tobin, Y.~Kr{\"a}henb{\"u}hl, L.~Kramer,
  \href{https://doi.org/10.1007/BF00117427}{Nonequilibrium theory of dirty,
  current-carrying superconductors: phase-slip oscillators in narrow filaments
  near tc}, J. Low Temp. Phys. 42~(5) (1981) 459--501.
\newline\urlprefix\url{https://doi.org/10.1007/BF00117427}

\bibitem{Kopnin2001-ip}
N.~B. Kopnin,
  \href{https://play.google.com/store/books/details?id=fYZb3D1V8rwC}{Theory of
  Nonequilibrium Superconductivity}, Clarendon Press, 2001.
\newline\urlprefix\url{https://play.google.com/store/books/details?id=fYZb3D1V8rwC}

\bibitem{Aranson2002-so}
I.~S. Aranson, L.~Kramer,
  \href{https://link.aps.org/doi/10.1103/RevModPhys.74.99}{The world of the
  complex {Ginzburg-Landau} equation}, Rev. Mod. Phys. 74~(1) (2002) 99--143.
\newline\urlprefix\url{https://link.aps.org/doi/10.1103/RevModPhys.74.99}

\bibitem{Blatter1994-mq}
G.~Blatter, M.~V. Feigel'man, V.~B. Geshkenbein, A.~I. Larkin, V.~M. Vinokur,
  \href{https://link.aps.org/doi/10.1103/RevModPhys.66.1125}{Vortices in
  high-temperature superconductors}, Rev. Mod. Phys. 66~(4) (1994) 1125--1388.
\newline\urlprefix\url{https://link.aps.org/doi/10.1103/RevModPhys.66.1125}

\bibitem{Kwok2016-of}
W.-K. Kwok, U.~Welp, A.~Glatz, A.~E. Koshelev, K.~J. Kihlstrom, G.~W. Crabtree,
  \href{http://dx.doi.org/10.1088/0034-4885/79/11/116501}{Vortices in
  high-performance high-temperature superconductors}, Rep. Prog. Phys. 79~(11)
  (2016) 116501.
\newline\urlprefix\url{http://dx.doi.org/10.1088/0034-4885/79/11/116501}

\bibitem{Alstrom2011-bc}
T.~S. Alstr{\o}m, M.~P. S{\o}rensen, N.~F. Pedersen, S.~Madsen,
  \href{https://doi.org/10.1007/s10440-010-9580-8}{Magnetic flux lines in
  complex geometry {Type-II} superconductors studied by the time dependent
  {Ginzburg-Landau} equation}, Acta Appl. Math. 115~(1) (2011) 63--74.
\newline\urlprefix\url{https://doi.org/10.1007/s10440-010-9580-8}

\bibitem{Oripov2020-dq}
B.~Oripov, S.~M. Anlage,
  \href{http://dx.doi.org/10.1103/PhysRevE.101.033306}{Time-dependent
  {Ginzburg-Landau} treatment of rf magnetic vortices in superconductors:
  Vortex semiloops in a spatially nonuniform magnetic field}, Phys Rev E
  101~(3-1) (2020) 033306.
\newline\urlprefix\url{http://dx.doi.org/10.1103/PhysRevE.101.033306}

\bibitem{Peng2017-zt}
L.~Peng, C.~Cai, \href{https://doi.org/10.1007/s10909-017-1769-z}{Finite
  element treatment of vortex states in {3D} cubic superconductors in a tilted
  magnetic field}, J. Low Temp. Phys. 188~(1) (2017) 39--48.
\newline\urlprefix\url{https://doi.org/10.1007/s10909-017-1769-z}

\bibitem{Machida1993-zm}
M.~Machida, H.~Kaburaki,
  \href{http://dx.doi.org/10.1103/PhysRevLett.71.3206}{Direct simulation of the
  time-dependent {Ginzburg-Landau} equation for {type-II} superconducting thin
  film: Vortex dynamics and {V-I} characteristics}, Phys. Rev. Lett. 71~(19)
  (1993) 3206--3209.
\newline\urlprefix\url{http://dx.doi.org/10.1103/PhysRevLett.71.3206}

\bibitem{Clem2011-ji}
J.~R. Clem, K.~K. Berggren,
  \href{https://link.aps.org/doi/10.1103/PhysRevB.84.174510}{Geometry-dependent
  critical currents in superconducting nanocircuits}, Phys. Rev. B Condens.
  Matter 84~(17) (2011) 174510.
\newline\urlprefix\url{https://link.aps.org/doi/10.1103/PhysRevB.84.174510}

\bibitem{Clem2012-og}
J.~R. Clem, Y.~Mawatari, G.~R. Berdiyorov, F.~M. Peeters,
  \href{https://link.aps.org/doi/10.1103/PhysRevB.85.144511}{Predicted
  field-dependent increase of critical currents in asymmetric superconducting
  nanocircuits}, Phys. Rev. B Condens. Matter 85~(14) (2012) 144511.
\newline\urlprefix\url{https://link.aps.org/doi/10.1103/PhysRevB.85.144511}

\bibitem{Berdiyorov2012-rn}
G.~R. Berdiyorov, M.~V. Milo{\v s}evi{\'c}, F.~M. Peeters,
  \href{https://doi.org/10.1063/1.4731627}{Spatially dependent sensitivity of
  superconducting meanders as single-photon detectors}, Appl. Phys. Lett.
  100~(26) (2012) 262603.
\newline\urlprefix\url{https://doi.org/10.1063/1.4731627}

\bibitem{Sardella2006-hi}
E.~Sardella, A.~L. Malvezzi, P.~N. Lisboa-Filho, W.~A. Ortiz,
  \href{https://link.aps.org/doi/10.1103/PhysRevB.74.014512}{Temperature-dependent
  vortex motion in a square mesoscopic superconducting cylinder:
  {Ginzburg-Landau} calculations}, Phys. Rev. B Condens. Matter 74~(1) (2006)
  014512.
\newline\urlprefix\url{https://link.aps.org/doi/10.1103/PhysRevB.74.014512}

\bibitem{Blair2018-og}
A.~I. Blair, D.~P. Hampshire,
  \href{http://dx.doi.org/10.1109/TASC.2018.2790985}{{Time-Dependent}
  {Ginzburg--Landau} simulations of the critical current in superconducting
  films and junctions in magnetic fields}, IEEE Trans. Appl. Supercond. 28~(4)
  (2018) 1--5.
\newline\urlprefix\url{http://dx.doi.org/10.1109/TASC.2018.2790985}

\bibitem{Jelic2016-ww}
{\v Z}.~L. Jeli{\'c}, M.~V. Milo{\v s}evi{\'c}, A.~V. Silhanek,
  \href{http://dx.doi.org/10.1038/srep35687}{Velocimetry of superconducting
  vortices based on stroboscopic resonances}, Sci. Rep. 6 (2016) 35687.
\newline\urlprefix\url{http://dx.doi.org/10.1038/srep35687}

\bibitem{Stosic2018-gb}
D.~Stosic, \href{https://repositorio.ufpe.br/handle/123456789/31433}{High
  performance {Ginzburg-Landau} simulations of superconductivity}, Ph.D.
  thesis, Universidade Federal de Pernambuco (2018).
\newline\urlprefix\url{https://repositorio.ufpe.br/handle/123456789/31433}

\bibitem{Winiecki2002-ka}
T.~Winiecki, C.~S. Adams,
  \href{https://www.sciencedirect.com/science/article/pii/S0021999102970476}{A
  fast {Semi-Implicit} {Finite-Difference} method for the {TDGL} equations}, J.
  Comput. Phys. 179~(1) (2002) 127--139.
\newline\urlprefix\url{https://www.sciencedirect.com/science/article/pii/S0021999102970476}

\bibitem{Hernandez2008-mi}
A.~D. Hern{\'a}ndez, D.~Dom{\'\i}nguez,
  \href{https://link.aps.org/doi/10.1103/PhysRevB.77.224505}{Dissipation spots
  generated by vortex nucleation points in mesoscopic superconductors driven by
  microwave magnetic fields}, Phys. Rev. B Condens. Matter 77~(22) (2008)
  224505.
\newline\urlprefix\url{https://link.aps.org/doi/10.1103/PhysRevB.77.224505}

\bibitem{Bezuglyj2022-cm}
A.~I. Bezuglyj, V.~A. Shklovskij, B.~Budinsk{\'a}, B.~Aichner, V.~M. Bevz,
  M.~Y. Mikhailov, D.~Y. Vodolazov, W.~Lang, O.~V. Dobrovolskiy,
  \href{https://link.aps.org/doi/10.1103/PhysRevB.105.214507}{Vortex jets
  generated by edge defects in current-carrying superconductor thin strips},
  Phys. Rev. B Condens. Matter 105~(21) (2022) 214507.
\newline\urlprefix\url{https://link.aps.org/doi/10.1103/PhysRevB.105.214507}

\bibitem{Al_Luhaibi2022-cl}
A.~Al~Luhaibi, A.~Glatz, J.~B. Ketterson,
  \href{https://link.aps.org/doi/10.1103/PhysRevB.106.224516}{Driven responses
  of periodically patterned superconducting films}, Phys. Rev. B Condens.
  Matter 106~(22) (2022) 224516.
\newline\urlprefix\url{https://link.aps.org/doi/10.1103/PhysRevB.106.224516}

\bibitem{Jonsson2022-xe}
M.~J{\"o}nsson,
  \href{http://kth.diva-portal.org/smash/record.jsf?pid=diva2%3A1657766&dswid=-9243}{Theory
  for superconducting few-photon detectors}, Ph.D. thesis, KTH Royal Institute
  of Technology (2022).
\newline\urlprefix\url{http://kth.diva-portal.org/smash/record.jsf?pid=diva2%3A1657766&dswid=-9243}

\bibitem{Jonsson2022-mb}
M.~J{\"o}nsson, R.~Vedin, S.~Gyger, J.~A. Sutton, S.~Steinhauer, V.~Zwiller,
  M.~Wallin, J.~Lidmar,
  \href{https://link.aps.org/doi/10.1103/PhysRevApplied.17.064046}{Current
  crowding in nanoscale superconductors within the {Ginzburg-Landau} model},
  Phys. Rev. Applied 17~(6) (2022) 064046.
\newline\urlprefix\url{https://link.aps.org/doi/10.1103/PhysRevApplied.17.064046}

\bibitem{Gropp1996-uw}
W.~D. Gropp, H.~G. Kaper, G.~K. Leaf, D.~M. Levine, M.~Palumbo, V.~M. Vinokur,
  \href{https://www.sciencedirect.com/science/article/pii/S0021999196900224}{Numerical
  simulation of vortex dynamics in {Type-II} superconductors}, J. Comput. Phys.
  123~(2) (1996) 254--266.
\newline\urlprefix\url{https://www.sciencedirect.com/science/article/pii/S0021999196900224}

\bibitem{Du1998-kt}
Q.~Du, R.~A. Nicolaides, X.~Wu,
  \href{http://www.jstor.org/stable/2587121}{Analysis and convergence of a
  covolume approximation of the {Ginzburg-Landau} model of superconductivity},
  SIAM J. Numer. Anal. 35~(3) (1998) 1049--1072.
\newline\urlprefix\url{http://www.jstor.org/stable/2587121}

\bibitem{Sadovskyy2015-ha}
I.~A. Sadovskyy, A.~E. Koshelev, C.~L. Phillips, D.~A. Karpeyev, A.~Glatz,
  \href{https://www.sciencedirect.com/science/article/pii/S0021999115002284}{Stable
  large-scale solver for {Ginzburg--Landau} equations for superconductors}, J.
  Comput. Phys. 294 (2015) 639--654.
\newline\urlprefix\url{https://www.sciencedirect.com/science/article/pii/S0021999115002284}

\bibitem{Gurevich1987-sv}
A.~V. Gurevich, R.~G. Mints,
  \href{https://link.aps.org/doi/10.1103/RevModPhys.59.941}{Self-heating in
  normal metals and superconductors}, Rev. Mod. Phys. 59~(4) (1987) 941--999.
\newline\urlprefix\url{https://link.aps.org/doi/10.1103/RevModPhys.59.941}

\bibitem{Zotova2012-nc}
A.~N. Zotova, D.~Y. Vodolazov,
  \href{https://link.aps.org/doi/10.1103/PhysRevB.85.024509}{Photon detection
  by current-carrying superconducting film: A time-dependent {Ginzburg-Landau}
  approach}, Phys. Rev. B Condens. Matter 85~(2) (2012) 024509.
\newline\urlprefix\url{https://link.aps.org/doi/10.1103/PhysRevB.85.024509}

\bibitem{Jing2018-qc}
Z.~Jing, H.~Yong, Y.~Zhou,
  \href{https://iopscience.iop.org/article/10.1088/1361-6668/aab3be/meta}{Thermal
  coupling effect on the vortex dynamics of superconducting thin films:
  time-dependent {Ginzburg--Landau} simulations}, Supercond. Sci. Technol.
  31~(5) (2018) 055007.
\newline\urlprefix\url{https://iopscience.iop.org/article/10.1088/1361-6668/aab3be/meta}

\bibitem{Du1998-rb}
Q.~Du, \href{http://www.jstor.org/stable/2585166}{Discrete gauge invariant
  approximations of a time dependent {Ginzburg-Landau} model of
  superconductivity}, Math. Comput. 67~(223) (1998) 965--986.
\newline\urlprefix\url{http://www.jstor.org/stable/2585166}

\bibitem{Li2005-gv}
X.~S. Li, \href{https://doi.org/10.1145/1089014.1089017}{An overview of
  {SuperLU}: Algorithms, implementation, and user interface}, ACM Trans. Math.
  Softw. 31~(3) (2005) 302--325.
\newline\urlprefix\url{https://doi.org/10.1145/1089014.1089017}

\bibitem{Harris2020-xv}
C.~R. Harris, K.~J. Millman, S.~J. van~der Walt, R.~Gommers, P.~Virtanen,
  D.~Cournapeau, E.~Wieser, J.~Taylor, S.~Berg, N.~J. Smith, R.~Kern, M.~Picus,
  S.~Hoyer, M.~H. van Kerkwijk, M.~Brett, A.~Haldane, J.~F. Del~R{\'\i}o,
  M.~Wiebe, P.~Peterson, P.~G{\'e}rard-Marchant, K.~Sheppard, T.~Reddy,
  W.~Weckesser, H.~Abbasi, C.~Gohlke, T.~E. Oliphant, Array programming with
  {NumPy}, Nature 585~(7825) (2020) 357--362.

\bibitem{Virtanen2020-zz}
P.~Virtanen, R.~Gommers, T.~E. Oliphant, M.~Haberland, T.~Reddy, D.~Cournapeau,
  E.~Burovski, P.~Peterson, W.~Weckesser, J.~Bright, S.~J. van~der Walt,
  M.~Brett, J.~Wilson, K.~J. Millman, N.~Mayorov, A.~R.~J. Nelson, E.~Jones,
  R.~Kern, E.~Larson, C.~J. Carey, {\.I}.~Polat, Y.~Feng, E.~W. Moore,
  J.~VanderPlas, D.~Laxalde, J.~Perktold, R.~Cimrman, I.~Henriksen, E.~A.
  Quintero, C.~R. Harris, A.~M. Archibald, A.~H. Ribeiro, F.~Pedregosa, P.~van
  Mulbregt, {SciPy 1.0 Contributors}, {SciPy} 1.0: fundamental algorithms for
  scientific computing in python, Nat. Methods 17~(3) (2020) 261--272.

\bibitem{Hunter2007-il}
{Hunter}, Matplotlib: A {2D} graphics environment, Computing in Science
  Engineering 9 (2007) 90--95.

\bibitem{Collette2013-rq}
A.~Collette,
  \href{https://play.google.com/store/books/details?id=lRCMAQAAQBAJ}{Python and
  {HDF5}: Unlocking Scientific Data}, ``O'Reilly Media, Inc.'', 2013.
\newline\urlprefix\url{https://play.google.com/store/books/details?id=lRCMAQAAQBAJ}

\bibitem{Grecco}
H.~Grecco, \href{https://pint.readthedocs.io/en/stable/index.html}{Pint online
  documentation}, last accessed: 2022-03-16.
\newline\urlprefix\url{https://pint.readthedocs.io/en/stable/index.html}

\bibitem{Klockner}
A.~Kl\"ockner, \href{https://documen.tician.de/meshpy/}{Meshpy online
  documentation}, last accessed: 2022-03-16.
\newline\urlprefix\url{https://documen.tician.de/meshpy/}

\bibitem{Shewchuk1996-va}
J.~R. Shewchuk, Triangle: Engineering a {2D} quality mesh generator and
  delaunay triangulator, in: Applied Computational Geometry Towards Geometric
  Engineering, Springer Berlin Heidelberg, 1996, pp. 203--222.

\bibitem{Bishop-Van_Horn2022-sy}
L.~Bishop-Van~Horn, K.~A. Moler,
  \href{https://www.sciencedirect.com/science/article/pii/S0010465522001837}{{SuperScreen}:
  An open-source package for simulating the magnetic response of
  two-dimensional superconducting devices}, Comput. Phys. Commun. 280 (2022)
  108464.
\newline\urlprefix\url{https://www.sciencedirect.com/science/article/pii/S0010465522001837}

\bibitem{Cao2018-rx}
Y.~Cao, V.~Fatemi, S.~Fang, K.~Watanabe, T.~Taniguchi, E.~Kaxiras,
  P.~Jarillo-Herrero,
  \href{http://dx.doi.org/10.1038/nature26160}{Unconventional superconductivity
  in magic-angle graphene superlattices}, Nature 556~(7699) (2018) 43--50.
\newline\urlprefix\url{http://dx.doi.org/10.1038/nature26160}

\bibitem{Fatemi2018-az}
V.~Fatemi, S.~Wu, Y.~Cao, L.~Bretheau, Q.~D. Gibson, K.~Watanabe, T.~Taniguchi,
  R.~J. Cava, P.~Jarillo-Herrero,
  \href{http://dx.doi.org/10.1126/science.aar4642}{Electrically tunable
  low-density superconductivity in a monolayer topological insulator}, Science
  362~(6417) (2018) 926--929.
\newline\urlprefix\url{http://dx.doi.org/10.1126/science.aar4642}

\bibitem{Park2021-pk}
J.~M. Park, Y.~Cao, K.~Watanabe, T.~Taniguchi, P.~Jarillo-Herrero,
  \href{http://dx.doi.org/10.1038/s41586-021-03192-0}{Tunable strongly coupled
  superconductivity in magic-angle twisted trilayer graphene}, Nature
  590~(7845) (2021) 249--255.
\newline\urlprefix\url{http://dx.doi.org/10.1038/s41586-021-03192-0}

\bibitem{Park2022-pu}
J.~M. Park, Y.~Cao, L.-Q. Xia, S.~Sun, K.~Watanabe, T.~Taniguchi,
  P.~Jarillo-Herrero,
  \href{http://dx.doi.org/10.1038/s41563-022-01287-1}{Robust superconductivity
  in magic-angle multilayer graphene family}, Nat. Mater. 21~(8) (2022)
  877--883.
\newline\urlprefix\url{http://dx.doi.org/10.1038/s41563-022-01287-1}

\bibitem{Skocpol1974-sc}
W.~J. Skocpol, M.~R. Beasley, M.~Tinkham,
  \href{https://doi.org/10.1007/BF00655865}{Phase-slip centers and
  nonequilibrium processes in superconducting tin microbridges}, J. Low Temp.
  Phys. 16~(1) (1974) 145--167.
\newblock \href {https://doi.org/10.1007/BF00655865}
  {\path{doi:10.1007/BF00655865}}.
\newline\urlprefix\url{https://doi.org/10.1007/BF00655865}

\bibitem{Ivlev1984-ct}
B.~I. Ivlev, N.~B. Kopnin,
  \href{https://doi.org/10.1080/00018738400101641}{Electric currents and
  resistive states in thin superconductors}, Adv. Phys. 33~(1) (1984) 47--114.
\newline\urlprefix\url{https://doi.org/10.1080/00018738400101641}

\bibitem{Sivakov2003-ss}
A.~G. Sivakov, A.~M. Glukhov, A.~N. Omelyanchouk, Y.~Koval, P.~M{\"u}ller,
  A.~V. Ustinov,
  \href{http://dx.doi.org/10.1103/PhysRevLett.91.267001}{Josephson behavior of
  phase-slip lines in wide superconducting strips}, Phys. Rev. Lett. 91~(26 Pt
  1) (2003) 267001.
\newblock \href {https://doi.org/10.1103/PhysRevLett.91.267001}
  {\path{doi:10.1103/PhysRevLett.91.267001}}.
\newline\urlprefix\url{http://dx.doi.org/10.1103/PhysRevLett.91.267001}

\bibitem{Vijay2009-is}
R.~Vijay, J.~D. Sau, M.~L. Cohen, I.~Siddiqi,
  \href{http://dx.doi.org/10.1103/PhysRevLett.103.087003}{Optimizing
  anharmonicity in nanoscale weak link josephson junction oscillators}, Phys.
  Rev. Lett. 103~(8) (2009) 087003.
\newblock \href {https://doi.org/10.1103/PhysRevLett.103.087003}
  {\path{doi:10.1103/PhysRevLett.103.087003}}.
\newline\urlprefix\url{http://dx.doi.org/10.1103/PhysRevLett.103.087003}

\bibitem{Dayem1967-ev}
A.~H. Dayem, J.~J. Wiegand,
  \href{https://link.aps.org/doi/10.1103/PhysRev.155.419}{Behavior of
  {Thin-Film} superconducting bridges in a microwave field}, Phys. Rev. 155~(2)
  (1967) 419--428.
\newline\urlprefix\url{https://link.aps.org/doi/10.1103/PhysRev.155.419}

\bibitem{Hasselbach2002-ii}
K.~Hasselbach, D.~Mailly, J.~R. Kirtley,
  \href{https://doi.org/10.1063/1.1448864}{Micro-superconducting quantum
  interference device characteristics}, J. Appl. Phys. 91~(7) (2002)
  4432--4437.
\newline\urlprefix\url{https://doi.org/10.1063/1.1448864}

\bibitem{Vasyukov2013-qh}
D.~Vasyukov, Y.~Anahory, L.~Embon, D.~Halbertal, J.~Cuppens, L.~Neeman,
  A.~Finkler, Y.~Segev, Y.~Myasoedov, M.~L. Rappaport, M.~E. Huber, E.~Zeldov,
  \href{http://dx.doi.org/10.1038/nnano.2013.169}{A scanning superconducting
  quantum interference device with single electron spin sensitivity}, Nat.
  Nanotechnol. 8~(9) (2013) 639--644.
\newline\urlprefix\url{http://dx.doi.org/10.1038/nnano.2013.169}

\bibitem{Nulens2022-iq}
L.~Nulens, H.~Dausy, M.~J. Wyszy{\'n}ski, B.~Raes, M.~J. Van~Bael, M.~V.
  Milo{\v s}evi{\'c}, J.~Van~de Vondel,
  \href{https://link.aps.org/doi/10.1103/PhysRevB.106.134518}{Metastable states
  and hidden phase slips in nanobridge {SQUIDs}}, Phys. Rev. B Condens. Matter
  106~(13) (2022) 134518.
\newline\urlprefix\url{https://link.aps.org/doi/10.1103/PhysRevB.106.134518}

\bibitem{Wyss2022-us}
M.~Wyss, K.~Bagani, D.~Jetter, E.~Marchiori, A.~Vervelaki, B.~Gross,
  J.~Ridderbos, S.~Gliga, C.~Sch{\"o}nenberger, M.~Poggio,
  \href{https://link.aps.org/doi/10.1103/PhysRevApplied.17.034002}{Magnetic,
  thermal, and topographic imaging with a {Nanometer-Scale} {SQUID-On-Lever}
  scanning probe}, Phys. Rev. Applied 17~(3) (2022) 034002.
\newline\urlprefix\url{https://link.aps.org/doi/10.1103/PhysRevApplied.17.034002}

\bibitem{Anahory2014-xy}
Y.~Anahory, J.~Reiner, L.~Embon, D.~Halbertal, A.~Yakovenko, Y.~Myasoedov,
  M.~L. Rappaport, M.~E. Huber, E.~Zeldov,
  \href{http://dx.doi.org/10.1021/nl503022q}{Three-junction {SQUID-on-tip} with
  tunable in-plane and out-of-plane magnetic field sensitivity}, Nano Lett.
  14~(11) (2014) 6481--6487.
\newline\urlprefix\url{http://dx.doi.org/10.1021/nl503022q}

\bibitem{Uri2016-qg}
A.~Uri, A.~Y. Meltzer, Y.~Anahory, L.~Embon, E.~O. Lachman, D.~Halbertal,
  N.~Hr, Y.~Myasoedov, M.~E. Huber, A.~F. Young, E.~Zeldov,
  \href{http://dx.doi.org/10.1021/acs.nanolett.6b02841}{Electrically tunable
  multiterminal {SQUID-on-Tip}}, Nano Lett. 16~(11) (2016) 6910--6915.
\newline\urlprefix\url{http://dx.doi.org/10.1021/acs.nanolett.6b02841}

\bibitem{Khapaev2001-pw}
M.~M. Khapaev, A.~Y. Kidiyarova-Shevchenko, P.~Magnelind, M.~Y. Kupriyanov,
  {3D-MLSI}: software package for inductance calculation in multilayer
  superconducting integrated circuits, IEEE Trans. Appl. Supercond. 11~(1)
  (2001) 1090--1093.

\bibitem{Fourie2011-wl}
C.~J. Fourie, O.~Wetzstein, T.~Ortlepp, J.~Kunert, Three-dimensional
  multi-terminal superconductive integrated circuit inductance extraction,
  Supercond. Sci. Technol. 24~(12) (2011) 125015.

\bibitem{Holmvall2022-ps}
P.~Holmvall, N.~Wall~Wennerdal, M.~H{\aa}kansson, P.~Stadler, O.~Shevtsov,
  T.~L{\"o}fwander, M.~Fogelstr{\"o}m,
  \href{http://arxiv.org/abs/2205.15000}{{SuperConga}: an open-source framework
  for mesoscopic superconductivity} (May 2022).
\newblock \href {http://arxiv.org/abs/2205.15000} {\path{arXiv:2205.15000}}.
\newline\urlprefix\url{http://arxiv.org/abs/2205.15000}

\bibitem{Seja2022-aa}
K.~M. Seja, T.~L{\"o}fwander,
  \href{https://link.aps.org/doi/10.1103/PhysRevB.106.144511}{Finite element
  method for the quasiclassical theory of superconductivity}, Phys. Rev. B
  Condens. Matter 106~(14) (2022) 144511.
\newline\urlprefix\url{https://link.aps.org/doi/10.1103/PhysRevB.106.144511}

\bibitem{Polyak1964-gb}
B.~T. Polyak,
  \href{https://www.sciencedirect.com/science/article/pii/0041555364901375}{Some
  methods of speeding up the convergence of iteration methods}, USSR
  Computational Mathematics and Mathematical Physics 4~(5) (1964) 1--17.
\newline\urlprefix\url{https://www.sciencedirect.com/science/article/pii/0041555364901375}

\end{thebibliography}






% \end{thebibliography}

\end{document}